\documentclass[journal=nalefd,manuscript=article]{achemso}
\usepackage{chemformula}
\usepackage[T1]{fontenc} 
\usepackage{amssymb,amsmath,graphics,graphicx,float,braket}
\usepackage{epstopdf}
\usepackage{array,multirow,setspace}
\usepackage[autostyle]{csquotes}
\usepackage{dcolumn}
\usepackage{bm}
\usepackage{physics}\usepackage{subscript}
\usepackage{soul}
\usepackage{subfigure}
\usepackage[utf8x]{inputenc} 
\usepackage{verbatim}
\usepackage{comment}

\title{Can magneto-transport properties provide insight into the functional groups in semiconducting MXenes?}
\author{Namitha Anna Koshi}
\affiliation{Indo-Korea Science and Technology Center (IKST), Jakkur, Bengaluru 560065, India}
\author{Anup Kumar Mandia}
\affiliation{Department of Electrical Engineering, Indian Institute of Technology Bombay, Powai, Mumbai-400076, India}
\author{Bhaskaran Muralidharan}
\affiliation{Department of Electrical Engineering, Indian Institute of Technology Bombay, Powai, Mumbai-400076, India}
\author{Seung-Cheol Lee}
\affiliation{Electronic Materials Research Center, KIST, Seoul 136-791, South Korea}
\email{leesc@kist.re.kr}
\author{Satadeep Bhattacharjee}
\affiliation{Indo-Korea Science and Technology Center (IKST), Jakkur, Bengaluru 560065, India}
\email{s.bhattacharjee@ikst.res.in}


\begin{document}
\begin{tocentry}
\includegraphics[scale=0.45]{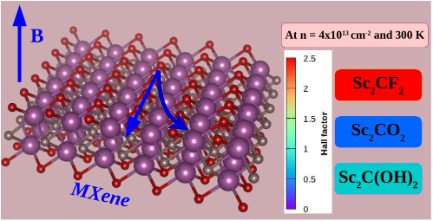}
\end{tocentry}
\begin{abstract}
The Hall scattering factor of \ch{Sc2CF2}, \ch{Sc2CO2} and \ch{Sc2C(OH)2} is calculated using Rode's iterative approach by solving the Boltzmann transport equation. This is carried out in conjunction with calculations based on density functional theory. The electrical transport in \ch{Sc2CF2}, \ch{Sc2CO2}, and \ch{Sc2C(OH)2} is modelled by accounting for both elastic (acoustic and piezoelectric) and inelastic (polar optical phonon) scattering. Polar optical phonon (POP) scattering is the most significant mechanism in these MXenes. We observe that there is a window of carrier concentration where Hall factor acts dramatically; \ch{Sc2CF2} obtains an incredible high value of 2.49 while \ch{Sc2CO2} achieves a very small value of approximately 0.5, and \ch{Sc2C(OH)2} achieves the so called ideal value of 1. We propose in this paper that such Hall factor behaviour has significant promise in the field of surface group identification in MXenes, an issue that has long baffled researchers.

\end{abstract}


\par
Following the discovery of graphene, two-dimensional (2D) materials are of great interest to researchers as they have a range of physical properties which makes them suitable for a variety of applications in energy storage, catalysis, high performance electronic devices and so on \cite{novoselov2004electric, mas20112d, akinwande2017review, novoselov2005two, neto2009electronic}. The reduced dimensionality of 2D materials promises to cater to the next generation electronic technologies and applications. MXenes, a family of 2D materials consisting of early transition metal carbides, nitrides and carbonitrides are known to possess unique characteristics \cite{barsoum2000mn+, barsoum2011elastic, naguib2011two, naguib2012two, naguib201425th, ghidiu2014synthesis}. They are formed by selective exfoliation of "A" from 3D layered MAX phases and sonication. MAX phases are precursors of MXene with space group $P6_3/mmc$ and chemical composition $M_{n+1}AX_{n}$ where M is an early transition metal, A is a group IIIA or IVA element and X is C and/or N with n=1-3. During the etching process, some functional groups (T) are left on the MXene surface so as to passivate the outer metal atom layer. From theoretical studies, it is known that bare MXenes ($M_{n+1}X_n$) are metallic and the electronic nature of functionalized MXenes ($M_{n+1}X_nT_x$) depends on the surface terminating species. Most of the functionalized MXenes retain the metallic nature of bare MXene and few MXenes like \ch{Ti2CO2}, \ch{Zr2CO2}, \ch{Hf2CO2}, \ch{Sc2CF2}, \ch{Sc2CO2} and \ch{Sc2C(OH)2} are reported to be semiconducting with appreciable band gap \cite{khazaei2013novel}. \ch{Sc2CO2} has an out-of-plane polarization $P_z$ with magnitude, an order higher than the 1T phase of \ch{MoS2} \cite{chandrasekaran2017ferroelectricity}. These 2D scandium carbides are not yet experimentally synthesized but have excellent properties from theoretical investigation. There are studies in which scandium carbide MXenes are used as a heterostructure in combination with the same base material or other 2D materials to develop a nanodevice for use in photonics, electronics and optoelectronics \cite{lee2015achieving, rehman2019van}. The semiconducting nature of scandium carbide MXenes make them promising candidate materials for future electronic and optical devices. \\

Intrinsic carrier mobility is a good measure of electrical transport and graphene has extremely high carrier mobility among 2D materials \cite{bolotin2008ultrahigh}. As graphene is gapless, it has low on-off ratio \cite{liao2010high}. Though extensive efforts are made to open a gap in graphene, it is apparently better to employ other 2D materials which has attributes like moderate band gap (ensures a high on-off ratio) and high intrinsic carrier mobility \cite{qiao2014high}. Most of the studied 2D semiconductors in their monolayer form have very low carrier mobility (< 100 cm$^{2}$/Vs) at room temperature \cite{li2019dimensional, cheng2019optimal, gunst2016first, li2014black} compared to bulk semiconductors like silicon and germanium. Therefore, it is essential to determine the factors that limit carrier transport in 2D semiconductors to design better electronic devices based on them. The carrier mobility is limited by the phonon interactions, which disturbs the potential experienced by electrons and its determination in principle involves the estimation of electron-phonon coupling in rigorous manner, methods for which have been developed in recent years~\cite{perturbo}. In the present work, we treat the acoustic phonons within deformation potential approach as it has produced quite satisfactory results for many MXene systems. \\

Magneto-transport has been widely studied in bulk semiconductors like \ch{Si} \cite{debye1954hall, kirnas1974concentration, putley1958electrical} and \ch{GaAs} \cite{stillman1970hall} and in 2D materials, graphene and \ch{WTe2} are reported to have large magnetoresistance (MR) \cite{matveev2018large, gopinadhan2013giant, wang2014classical, ali2014large}. We do not include spin-orbit coupling (SOC) in our \textit{ab-initio} calculations as it is more important for holes. Desai \textit{et al.}\cite{desai2021magnetotransport} demonstrated that to obtain accurate isotropic conductivity tensor of silicon for hole carriers, the effect of SOC has to be considered. The mismatch in MR anisotropy obtained in calculations is due to the errors in conductivity tensor determined without including SOC. Therefore, the key to accurate magneto-transport calculations for hole carriers is the inclusion of SOC and it has a minor effect on electron carriers. Studies on magneto-transport of materials using first principles are very scarce. Macheda \textit{et al.} investigated the magnetotransport in diamond \cite{macheda2018magnetotransport}, graphene \cite{macheda2020theory} and our recent work on \ch{Ti2CO2} \cite{mandia2022electrical} are the few existing reports. Also, the tranverse magnetoresistance of copper, bismuth and \ch{WP2} are determined by combining \textit{ab-initio} Fermi surfaces with Boltmann transport theory by Zhang \textit{et al.}~\cite{zhang2019magnetoresistance}. Magneto-transport in semiconductors was used primarily to quantify carrier concentration, which facilitates in the design and production of efficient electronic devices. We show that magneto-transport coefficients such as the Hall factor may be used as a guideline or tool to determine the functional groups in semiconducting MXenes such as metal-carbides. 
The electrical and magneto transport properties are calculated by the AMMCR code  \cite{mandia2021ammcr, mandia2019ab} developed by us. When no magnetic field is present, considering a spatial homogeneous system in the steady state (under an electric field E), the following equation is solved to obtain the single electron distribution function $f({\bf r},{\bf k},t)$ given by,
\begin{equation}
\frac{e\textbf{E}}{\hbar}\cdot\nabla _kf = \int [ s(k,k')f(1-f') -  s(k',k)\; f'(1-f)] dk' ,
\label{BTE1}
\end{equation}

where $s(k,k')$ represents the transition rate of an electron from a state $k$ to a state $k'$.  At lower electric fields, the distribution function is given by \cite{rode1970electron, rode1970electron1, rode1975low}

\begin{equation} 
f(k) = f_0[\epsilon(k)] +  g(k)cos\theta,
\label{dist_ft}
\end{equation}

where $f_0[\epsilon(k)]$ is the equilibrium distribution function given by $f_0[\epsilon(k)]=\frac{1}{e^{(\epsilon(k)-E_F)/k_BT}+1}$ $E_F$ is the Fermi-energy, \(g(k)\) is perturbation to the distribution function due to the presence of the applied electric field, and \(\cos\theta\) is the angle between applied electric field and \(k\). Here, we neglect higher order terms as mobility is calculated under low electric field conditions. It is required to calculate the perturbation in the distribution function \(g(k)\) for determining the low-field transport properties. Within the iterative formalism introduced by Rode, the perturbation to the distribution function \(g(k)\) is given by \cite{rode1970electron, rode1970electron1, rode1975low} 

\begin{equation}
g_{k,i+1} = \frac{S_i(g_k,i)- v(k)(\frac{\partial f}{\partial z})- \frac{eE}{\hbar}(\frac{\partial f}{\partial k}) } {S_o(k)} .
\label{pert}
\end{equation} 

where $S_{i}$ represents in-scattering rates  due to the inelastic processes  and $S_{o}$ represents the sum of out-scattering rates. $S_o=\frac{1}{\tau_{in}(k)}+\frac{1}{\tau_{el}(k)}$, where $\frac{1}{\tau_{el}(k)}$ is the sum of the momentum relaxation rates of all elastic scattering processes and  $\frac{1}{\tau_{in}(k)}$ is the momentum relaxation rate due to the in-elastic processes.

The expression for  $\tau_{el}(k)$, $S_{i}$ and $\frac{1}{\tau_{in}(k)}$ are given by the following equations
\begin{equation}
\frac{1}{\tau_{el}(k)} = \int (1 - X) s_{el}(k,k') dk'
\label{el}
\end{equation}

\begin{equation}
\ S_i(g_k,i) = \int X g_{k',i} [s_{in}(k',k)(1-f) + s_{in}(k,k')f ]dk'
\label{in_sc}
\end{equation}

\begin{equation}
\frac{1}{\tau_{in}(k)}=\int [s_{in}(k,k')(1 - f') + s_{in}(k',k)f']dk'
\label{out_sc}
\end{equation}

where X is the cosine of the angle between the initial and the final wave vectors, $s_{in}(k,k')$ and $s_{el}(k,k')$ represents transition rate of an electron from state $k$ to $k'$ due to inelastic and elastic scattering mechanisms respectively. Since, $S_{i}$ is function of g(k), thus equation \ref{pert} is to be calculated iteratively \cite{rode1975low}. In our previous work we have used the same procedure to calculate mobility for ZnSe \cite{mandia2019ab} and CdS\cite{mandia2021ammcr}. Drift mobility $\mu$ is then calculated by the following expression \cite{mandia2019ab} 

\begin{equation}
\mu = \frac{1}{2E} \frac{\int v(\epsilon) D_s(\epsilon) g(\epsilon) d\epsilon}  
{\int D_s(\epsilon) f( \epsilon )  d\epsilon },
\label{mobility}
\end{equation}
\quad

where $D_s(\epsilon)$ represents density of states. The carrier velocity is then calculated directly from the ab-initio band structure by using the form, $\ v(k) = \frac{1}{\hbar} \frac{\partial{\epsilon}}{\partial{k}}$ and the electrical conductivity can be calculated as $\sigma = \frac{n e \mu_e}{t_z}$. Here $n$ is the electron carrier concentration, $t_z$ is the thickness of the MXene layers along z-direction.

When we consider the magnetic field as well, the distribution function in this case is given by~\cite{rode1973theory}
\begin{equation} 
f(k) = f_0[\epsilon(k)] +  x g(k) + y h(k),
\label{dist_ft_bf}
\end{equation}

where h(k) represents perturbation in distribution function due to magnetic field, and y is direction cosine from $ \textbf{B} \times \textbf{E} $ to $\textbf{k}$, where $B$ is applied magnetic field. Substituting equation \ref{dist_ft_bf} in equation \ref{BTE1}, we get a pair of coupled equations that can be solved iteratively \cite{rode1973theory} 
  
\begin{equation}
g_{i+1}(k) = \frac{S_i(g_{i}(k))- \frac{eE}{\hbar}(\frac{\partial f_0}{\partial k}) + \beta S_i(h_{i}(k))} {S_o(k)(1 + \beta^2) } .
\label{pert_g}
\end{equation} 
 
\begin{equation}
h_{i+1}(k) = \frac{S_i(h_{i}(k))+ \beta \frac{eE}{\hbar}(\frac{\partial f_0}{\partial k}) - \beta S_i(g_{i}(k))} {S_o(k)(1 + \beta^2) } .
\label{pert_h}
\end{equation}

where $\beta = \frac{e v(k) B}{\hbar k S_o(k)}$.
The above expression shows that the perturbations to the distribution function due to the electric field ($g$) and magnetic field ($h$) are coupled to each other through the factor $\beta$ and the in scattering rates $S_i$. It should be highlighted that such a representation cannot be obtained using standard relaxation time approximation (RTA) and can only be seen using the current method.
The components of conductivity tensor in terms of perturbations are given by  
\begin{equation}
\sigma_{xx} =   \frac{e \int v(\epsilon) D_s(\epsilon) g(\epsilon) d\epsilon}{2E} 
\label{sigma_xx}
\end{equation}

\begin{equation}
\sigma_{xy} =   \frac{e \int v(\epsilon) D_s(\epsilon) h(\epsilon) d\epsilon}{2E} 
\label{sigma_xy}
\end{equation}

The Hall coefficient $R_{H}$, Hall mobility $\mu_H$ and Hall factor $r$ are respectively calculated by 
\begin{equation}
R_{H} =  \frac{\sigma_{xy}}{B(\sigma_{xx}\sigma_{yy} + \sigma_{xy}^2)} 
\label{hall_coeff}
\end{equation}

\begin{equation}
\mu_{H} =  \sigma_{xx}(0)|R_{H}|
\label{hall_mobility}
\end{equation}

\begin{equation}
r =  \frac{\mu_{H}}{\mu}
\label{hall_factor}
\end{equation}

where $\sigma_{xx}(0)$ is value of $\sigma_{xx}$ in the absence of the magnetic field.
We have already discussed three scattering mechanisms: scattering due to acoustic phonons, piezoelectric scattering, and scattering due to polar optical phonons. Ref. [\citenum{mandia2022electrical}] describes how to determine the scattering rates in detail.\\

In experiments, Hall measurements are carried out to estimate carrier concentration and drift mobility. A key quantity that characterize these measurements is the Hall scattering factor, r, which is taken as the ratio of Hall mobility to drift mobility. In general, this factor is assumed to be around unity and this assumption is based on compound semiconductors with quasi-parabolic bands \cite{macheda2020theory}. In such cases, r dependence on temperature and scattering times is weak. Here, we calculate Hall factor, r as
\begin{equation}
r=ne R_H=\frac{ne}{B}\frac{\sigma_{xy}}{(\sigma_{xx}\sigma_{yy}+\sigma_{xy}^2)}
\sim \frac{ne}{B}\frac{\sigma_{xy}}{\sigma_{xx}^2}
\label{factor}
\end{equation}
This equation can be rewritten as Equation \ref{r} by replacing the conductivity tensor components $\sigma_{xx}$ and $\sigma_{xy}$ with their expressions including pertubations to distribution function. The r dependence on temperature and scattering rates arises through these functions ($g(\varepsilon)$ and $h(\varepsilon)$).
\begin{equation}
r=\dfrac{n}{B}2E\frac{\int v(\epsilon)D_s(\varepsilon)h(\varepsilon)d\varepsilon}{[\int v(\epsilon)D_s(\varepsilon)g(\varepsilon)d\varepsilon]^2}
\label{r}
\end{equation}

We performed first principles calculations using density functional theory (DFT) implemented in plane wave code, Vienna ab-initio Simulation Package (VASP) \cite{kresse1996efficiency, kresse1996efficient}. The projector augmented wave (PAW) approach is employed for pseudopotentials \cite{blochl1994projector, kresse1999ultrasoft}. The generalized gradient approximation (GGA) with the Perdew-Burke-Ernzerhof (PBE) functional is adopted for exchange-correlation interactions \cite{perdew1996generalized}. The electron wavefunctions are expanded in plane waves with cut off energy of 500 eV. The conjugate gradient algorithm is used for structural optimization. The atomic positions are fully relaxed until the residual force on each atom is less than 0.01 eV/\AA{} and the energy convergence criterion is $10^{-6}$ eV. A vacuum of thickness 20 \AA{} along the z-direction is employed to avoid interactions between the neighboring layers and Brillouin zone integrations are performed with Gamma-centered k-mesh of 18$\times$18$\times$1 for structure optimizations. The DFT-D2 method is used for van der Waals correction \cite{grimme2010consistent}. The crystal structures are visualized using VESTA \cite{momma2008vesta}. Phonon spectra of \ch{Sc2CF2}, \ch{Sc2CO2} and \ch{Sc2C(OH)2} are calculated using VASP in combination with Phonopy software \cite{togo2015first}. Here, we employ a supercell of size 4$\times$4$\times$1 and a 4$\times$4$\times$1 k-mesh to determine the dynamical matrix. \\

The \ch{Sc2C} monolayer is modelled from \ch{Sc2AlC} bulk phase. 1T and 2H phases of \ch{Sc2C} are considered and the energetically more stable 1T phase is further used to study the effect of functionalization (F, O, OH). Depending on the position of functional groups on the bare MXene surface, there are four different configurations (I, II, III, IV). For \ch{Sc2CF2}, these configurations are given in Figure S1 and their total energies are compared for each functional group. Among these geometries, configuration II of \ch{Sc2CF2} and \ch{Sc2C(OH)2} have the lowest total energy.  The O functionalization results in a structure (configuration IV) different from that of F and OH. It is also reflected in the bond lengths and bond angles of \ch{Sc2CO2} which is mentioned by Kumar \textit{et al.}~\cite{kumar2016thermoelectric}. The optimized geometries for \ch{Sc2CF2}, \ch{Sc2CO2} and \ch{Sc2C(OH)2} are presented in Figure S1. In optimized structure of \ch{Sc2CO2} (Figure S1(f)), the oxygen atom on top lies in line with the scandium atom in the lower layer and the bottom oxygen atom is in line with carbon atom. For \ch{Sc2CF2} and \ch{Sc2C(OH)2}, the functional group on top lies in line with scandium atom in the lower layer and vice-versa. The symmetry group of \ch{Sc2CF2} and \ch{Sc2C(OH)2} is $P-3m_{1}$ (No. 164) whereas it is $P3m_{1}$ (No. 156) for \ch{Sc2CO2}. There is no inversion centre for \ch{Sc2CO2}. The layer thickness of \ch{Sc2CF2}, \ch{Sc2CO2} and \ch{Sc2C(OH)2} are 4.79, 3.85 and 6.91 \AA{} respectively, with \ch{Sc2CO2} exhibiting smallest layer thickness. \\

\begin{figure}
\subfigure[$ $]{\hspace{-1.5cm}\includegraphics[scale=0.32]{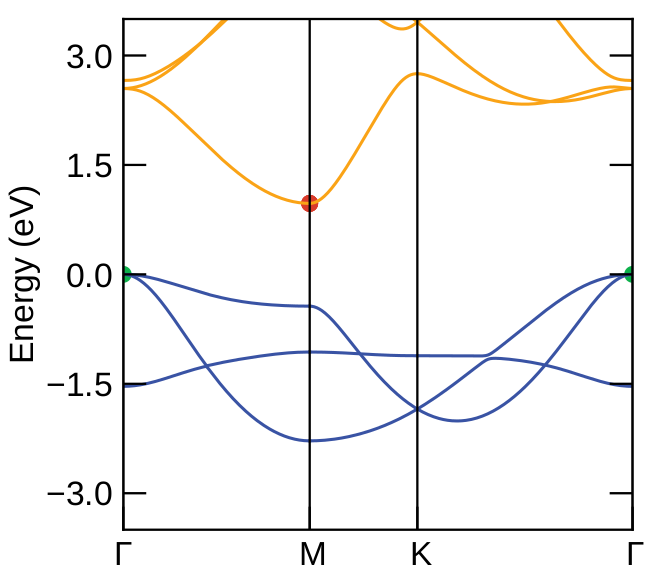}}
\subfigure[$ $]{\hspace{0cm}\includegraphics[trim=0mm 2mm 0mm 0mm,clip,scale=0.33]{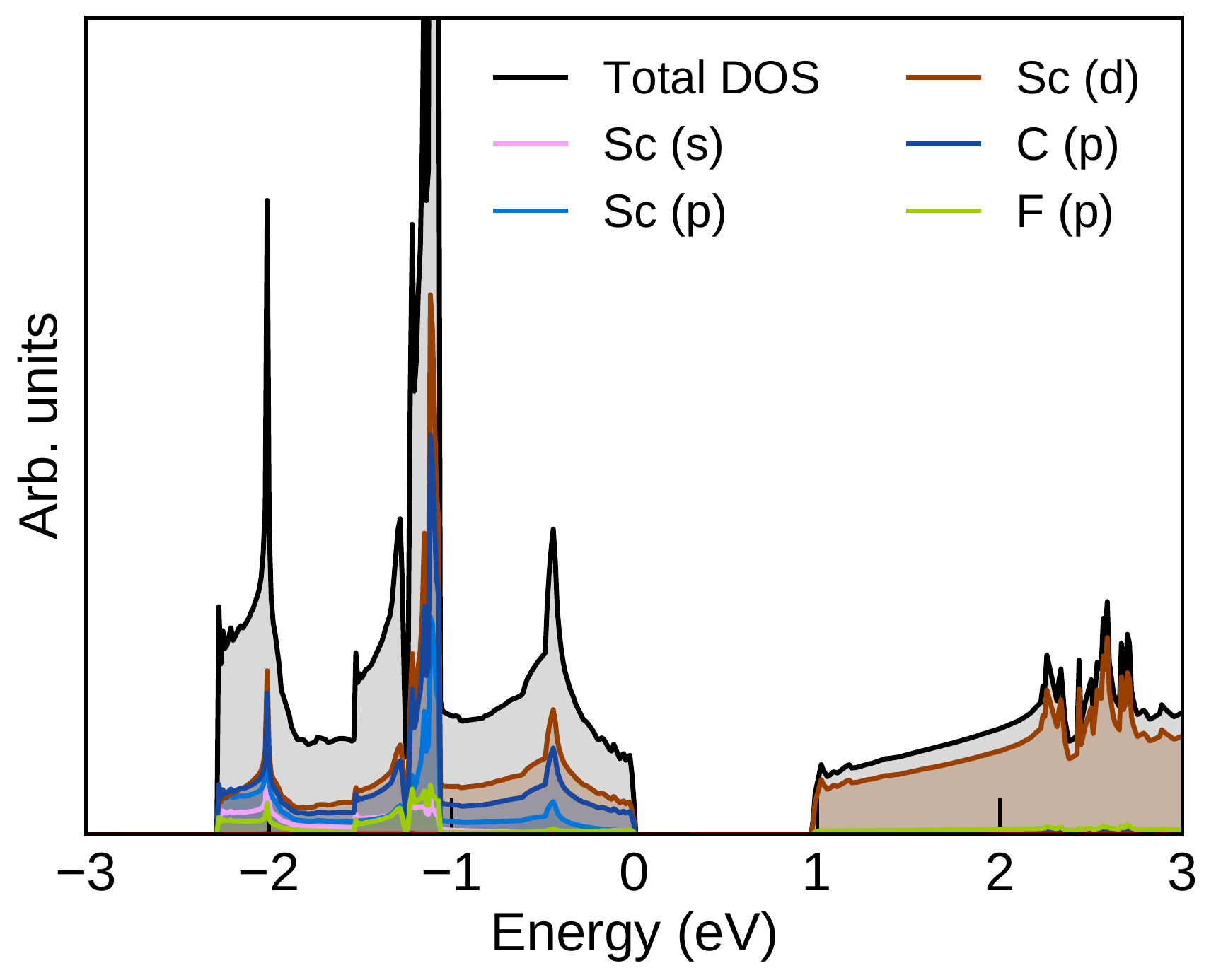}}
\subfigure[$ $]{\hspace{0cm}\includegraphics[scale=0.31]{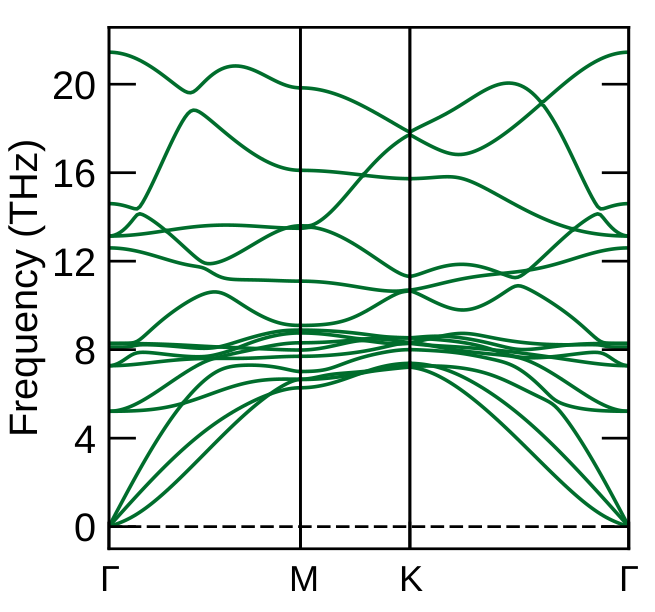}} \\
\subfigure[$ $]{\hspace{-1.5cm}\includegraphics[scale=0.33]{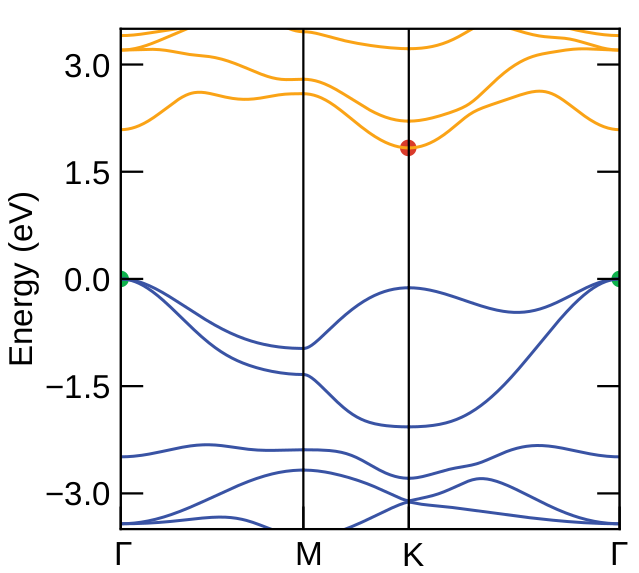}}
\subfigure[$ $]{\hspace{0cm}\includegraphics[trim=0mm 2mm 0mm 0mm,clip,scale=0.33]{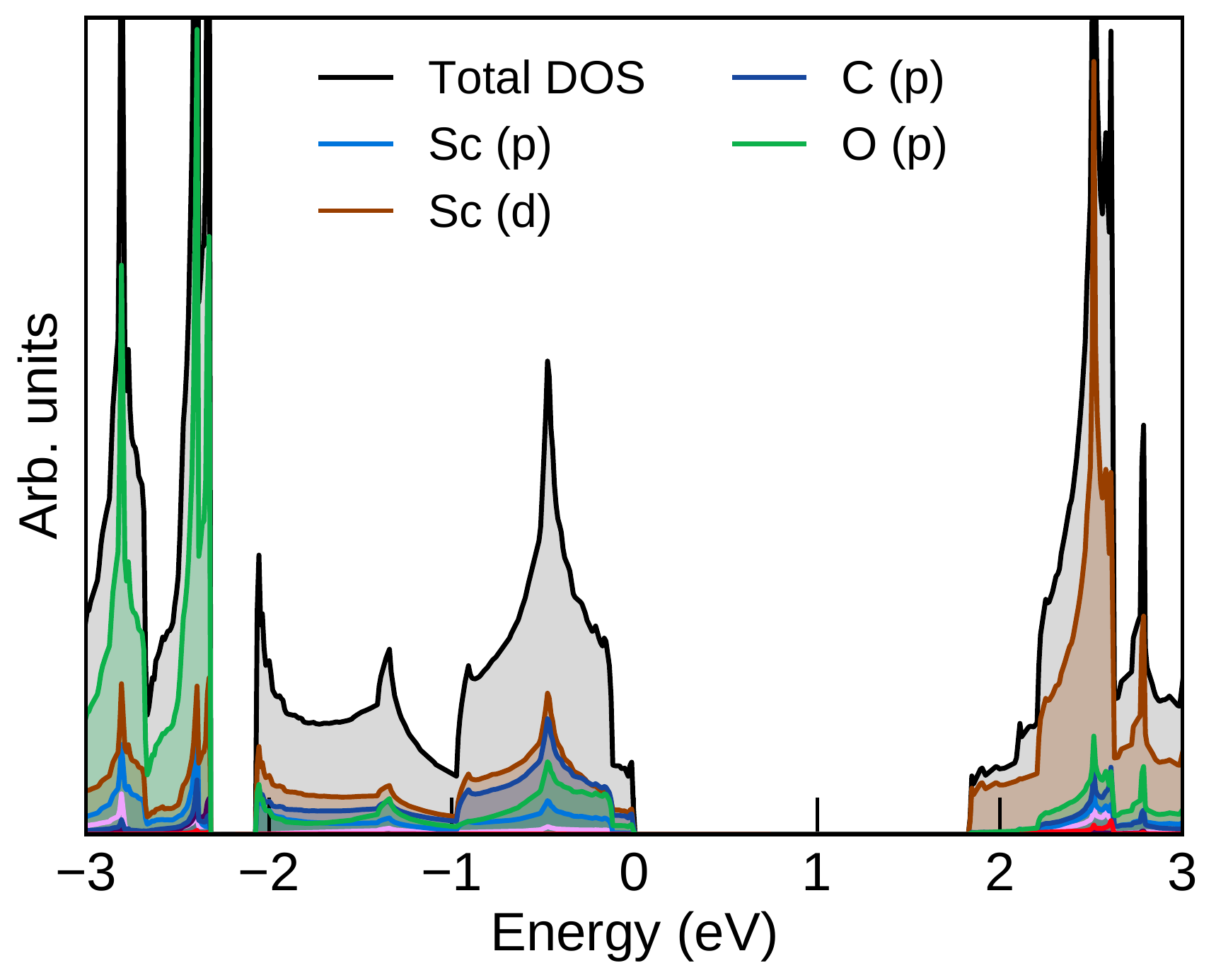}}
\subfigure[$ $]{\hspace{0cm}\includegraphics[scale=0.31]{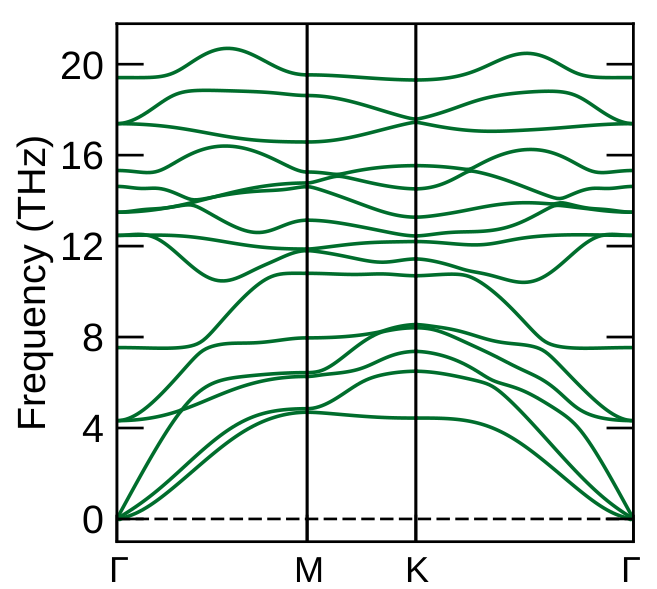}} \\
\subfigure[$ $]{\hspace{-1.2cm}\includegraphics[scale=0.32]{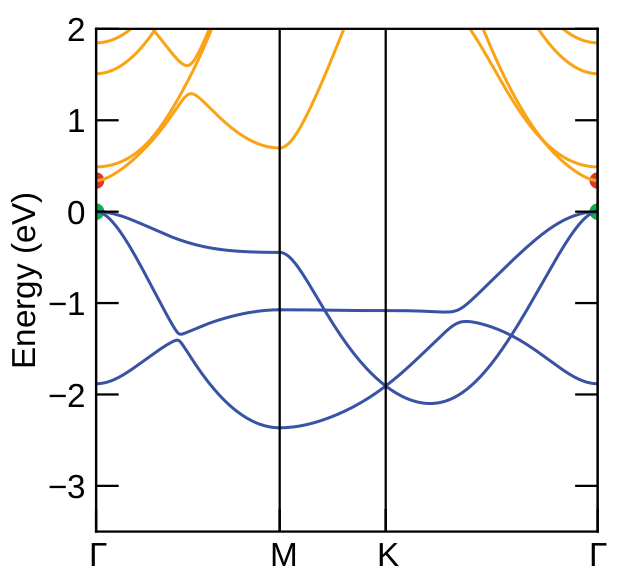}}
\subfigure[$ $]{\hspace{0cm}\includegraphics[trim=0mm 2mm 0mm 0mm,clip,scale=0.33]{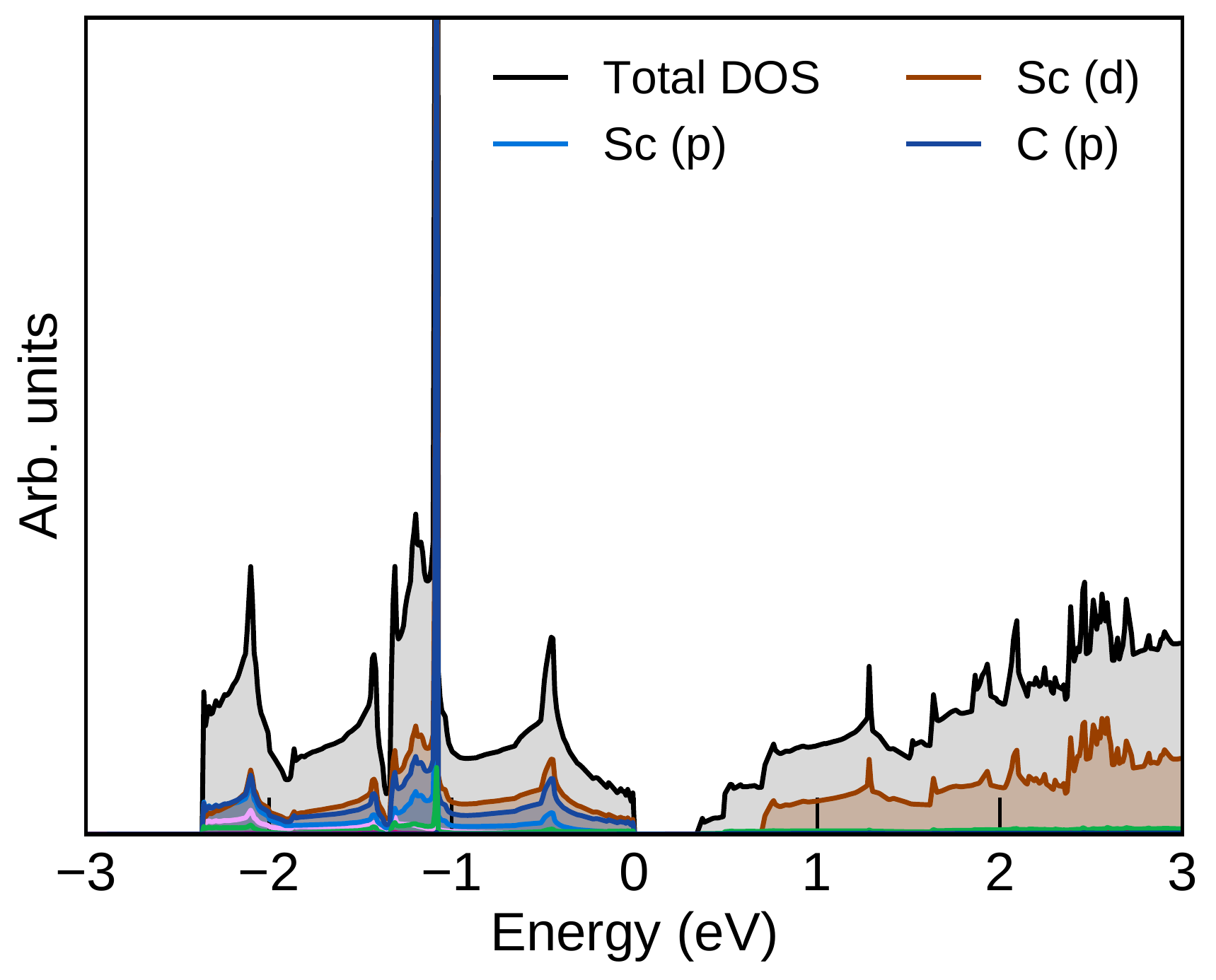}}
\subfigure[$ $]{\hspace{0cm}\includegraphics[scale=0.31]{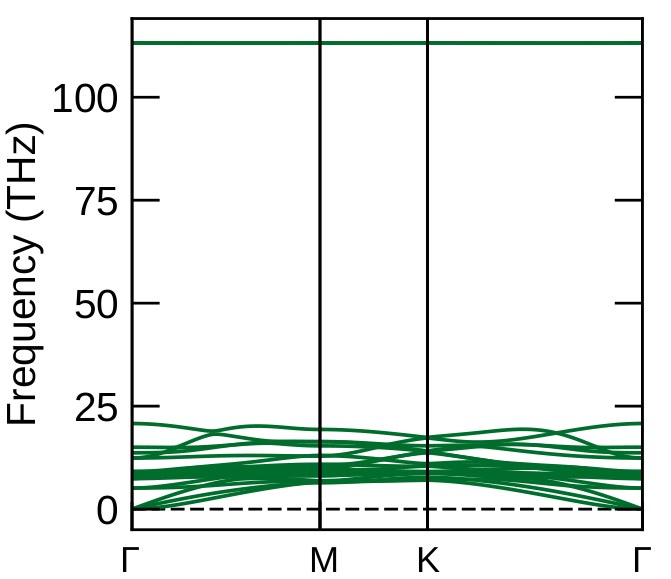}} 
 \caption{Electronic band structure, density of states and phonon band structure of \ch{Sc2CF2} (first row), \ch{Sc2CO2} (second row) and \ch{Sc2C(OH)2} (third row). The y (vertical) scale of DOS plots are multiplied by 2, 2 and 5 factors for F, O and OH functional groups respectively. In the electronic band structure, the VBM and CBM are denoted by green and red points respectively.}
 \label{figure 1}
\end{figure}
%

The electronic band structure and density of states (DOS) of three MXenes \ch{Sc2CF2}, \ch{Sc2CO2} and \ch{Sc2C(OH)2} are given in Figure \ref{figure 1}. They are semiconducting with indirect band gaps except \ch{Sc2C(OH)2}. Their band character also differs, which could be due to the distinct optimum structures for different functionalizations (from fat bands - not shown here). \ch{Sc2C(OH)2} has a direct band gap ($E_g$) of 0.34 eV with valence band maximum (VBM) and conduction band minimum (CBM) at $\Gamma$. The CBM is dominated by O and H bands whereas Sc has more contribution towards the VBM.  For \ch{Sc2CF2} and \ch{Sc2CO2}, the $E_g$ values are 0.97 and 1.83 eV respectively and CBM are formed by the Sc bands. Both of them have their VBM at $\Gamma$ and CBM at M and K respectively. These results are in agreement with reported ones \cite{khazaei2013novel, kumar2016thermoelectric, zha2015role}. The phonon dispersion spectra of \ch{Sc2CF2}, \ch{Sc2CO2} and \ch{Sc2C(OH)2} are also presented in Figure \ref{figure 1}. They do not have imaginary (negative) frequencies and hence these semiconducting MXenes are dynamically stable. For OH, there are high frequency optical branches above 100 THz. This feature is observed in other OH-functionalized MXenes and correspond to the stretching modes of OH bonds \cite{khazaei2015oh, hu2015vibrational}. It can also be used to characterize the level of OH functionalization in MXenes which after chemical exfoliation have different functional groups on the surface \cite{kumar2016thermoelectric}. \\ 

The \textit{ab-initio} parameters required to use the transport tool AMMCR \cite{mandia2021ammcr, mandia2019ab} are obtained from DFT calculations and these values are reported in Table-\ref{table1}. The piezoelectric and dielectric constants (both high and low frequency) are calculated using density functional perturbation theory (DFPT)~\cite{DFPT}. The acoustic deformation potential constant of CBM for electrons  and the elastic moduli under uniaxial strain are calculated along the out-of-plane (ZA), longitudinal (LA) and transverse (TA) directions using the method described by Xiong \textit{et al}~\cite{xiong2017functional}. For \ch{Sc2CF2}, the calculation of acoustic deformation potentials and elastic moduli are given in supporting information. \\

\begin{table} 
\caption{Material parameters used for the MXenes}
\begin{tabular}{ccccccccc}
Parameters  &  \ch{Sc2CF2} & \ch{Sc2CO2} & \ch{Sc2C(OH)2}   \\ \\
\hline \\
PZ constant, $e_{11} $ (C/cm)      &   $4.79 \times 10^{-17}$ & $42.35 \times 10^{-17}$ & $55.28 \times 10^{-17}$  \\
Acoustic deformation potentials, $D_{A} (eV):$       &    \\
$D_{A,LA} $  & 2.166 & 5.128 & 0.832   \\
$D_{A,TA} $  & 2.17  & 5.13 & 0.856 \\
$D_{A,ZA} $  & 6.842 & 0.326 & 3.396 \\
Elastic modulus, $C_{A} (N/m) $ & 		\\
$C_{A,LA}  $    & $243.08 $ & $251.30$ & $239.53$   \\
$C_{A,TA}  $    & $243.08 $ & $251.30$ & $239.53$ \\
$C_{A,ZA}  $    & $174.26 $ & $115.77$ & $355.62$   \\
Polar optical phonon frequency $\omega_{pop} (THz)$ &  \\
$\omega_{pop,LO} $  & 7.29 & 4.33  & 7.41 \\
$\omega_{pop,TO} $  & 7.29 & 4.33  & 7.41 \\
$\omega_{pop,HP} $  & 12.60 & 7.54 & 13.67  \\
High frequency dielectric constant, $ \kappa_{\infty}$  & 2.66 & 2.07 & 3.68  \\
Low frequency dielectric constant, $ \kappa_{0}$     &  7.43 & 3.33 & 9.19 \\

\end{tabular}
\label{table1}
\end{table}

%

\begin{figure}
    \subfigure[$ $]{\hspace{-1cm}\includegraphics[scale=0.25]{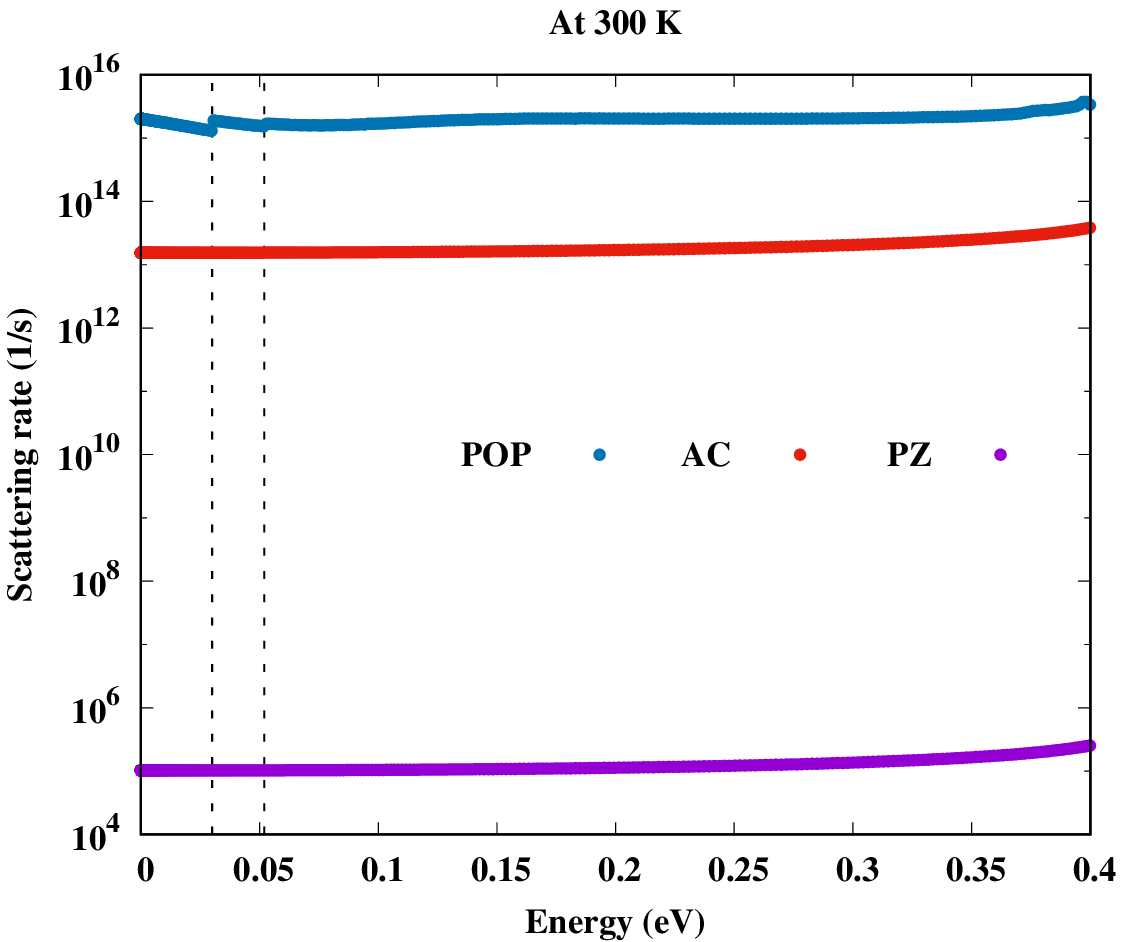}}
    \subfigure[$ $]{\hspace{-0.2cm}\includegraphics[trim=12mm 1mm 0mm 0mm,clip,scale=0.25]{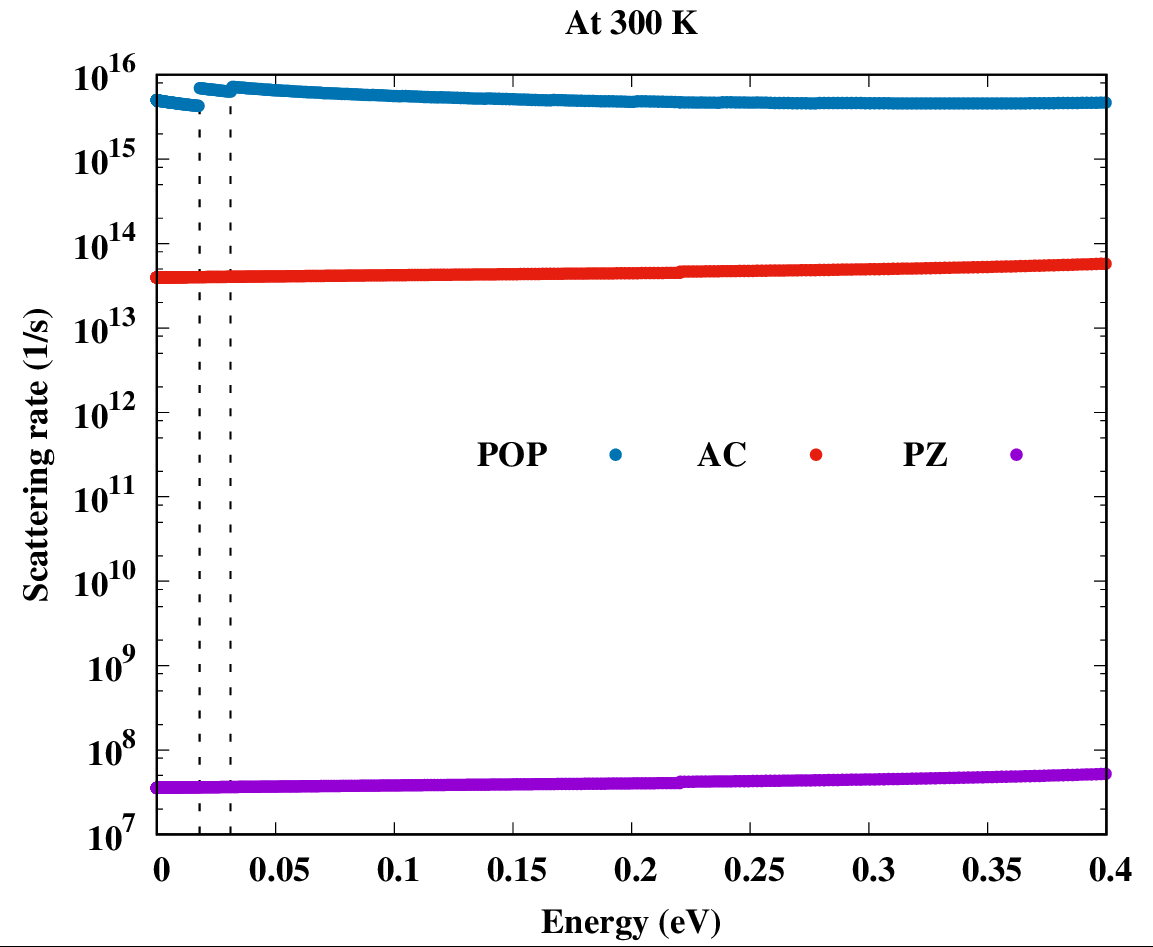}}
    \subfigure[$ $]{\hspace{-0.2cm}\includegraphics[trim=12mm 1mm 0mm 0mm,clip,scale=0.25]{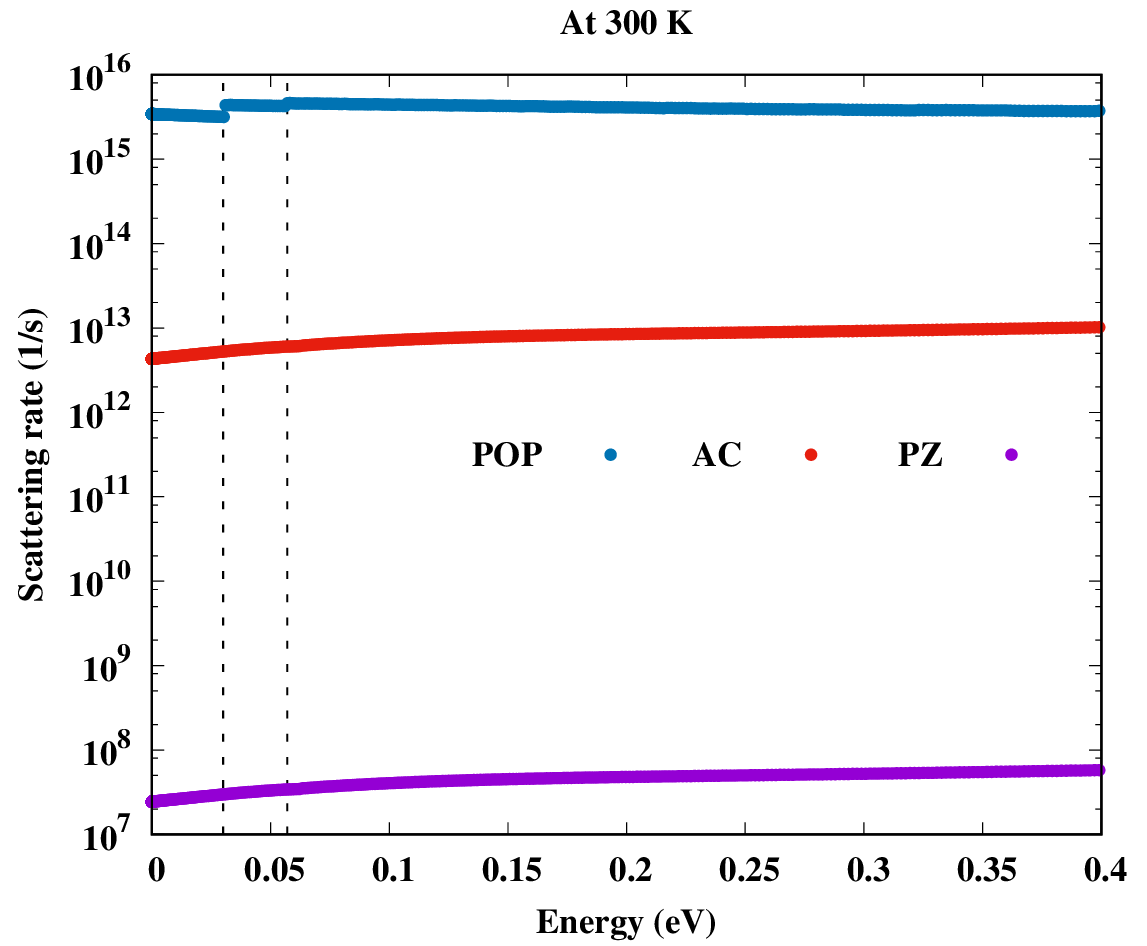}}
    \caption{Scattering rate versus energy for (a)\ch{Sc2CF2}, (b)\ch{Sc2CO2} and (c)\ch{Sc2C(OH)2}.}
    \label{scattering}
\end{figure}

Figure \ref{scattering} shows the scattering rates due to both acoustic and optical phonons for \ch{Sc2CF2}, \ch{Sc2CO2} and \ch{Sc2C(OH)2}. The out-of-plane phonons also contribute towards electron transport and are denoted by ZA/ZO modes. We find that for the three scandium carbide MXenes considered here, polar optical phonon (POP) interaction is the most dominant scattering mechanism, followed by deformation potential scattering and piezoelectric the least. For \ch{Sc2CF2} and \ch{Sc2C(OH)2}, it seen that the out-of-plane acoustic phonon called flexural phonon (ZA) has higher contribution when compared to the phonon scatterings induced by longitudinal (LA) and transverse (TA) modes. It is the opposite for \ch{Sc2CO2} with ZA phonon scattering weaker than LA and TA phonons. The contribution of LA, TA and ZA phonon modes towards scattering rate is shown in supporting information (Figure S3). \\

In these semiconductors, there is strong inelastic scattering as polar optical phonons play a major role. The black dashed lines in Figure \ref{scattering} corresponds to the energy of LO/TO and ZO phonon modes. When the electron energy is smaller than optical phonon energy, the interactions are dominated by the absorption process. As energy increases, the jumps in scattering rate represent the onset of phonon emission process. In these MXenes, the two lowest optical branches  (softer ones) are nonpolar modes/antiferroelectric, that is, the two Sc atoms (cations) have phonon dispersion eigen vectors pointing in opposite directions and so is the case of two T atoms (anions). Their mode effective charges calculated from Born effective charge have values close to zero. The next two branches are polar optical modes with the cations and anions vibrating in counterphase. In the case of \ch{Sc2CF2}, two Sc atoms vibrate in the same direction whereas the anions, C and F in the other direction. Here, the mode effective charge have significant values. We have calculated the scattering rates with both set of frequencies and it is observed that optical phonon interactions are dominant. The mobility in these 2D materials are limited by POP (due to the reciprocal nature of the relation) and it is represented as a function of temperature and electron concentration using the Matthiessen's rule
\begin{equation}
\frac{1}{\mu}=\frac{1}{\mu_{AC}}+\frac{1}{\mu_{PZ}}+\frac{1}{\mu_{POP}}
\label{Matt}
\end{equation}
where $\mu$, $\mu_{AC}$, $\mu_{PZ}$, and $\mu_{POP}$ denotes the total mobility, acoustic, piezoelectric, and polar optical contributions, respectively. Figure \ref{mobility} shows the variation of mobility as a function of temperature for \ch{Sc2CF2}, \ch{Sc2CO2} and \ch{Sc2C(OH)2}. For a particular electron concentration, the mobility decreases with increasing temperature. The mobility of \ch{Sc2CF2} decreases monotonously with increasing electron concentration upto 5$\times$10$^{12}$ cm$^{-2}$ and the curves corresponding to different doping concentrations coincide at high temperatures (exceeding 500 K). Beyond this carrier concentration, the change in mobility with increasing electron concentration is irregular. \ch{Sc2C(OH)2} shows similar behaviour but for \ch{Sc2CO2}, these limits are different. \\ 
\begin{figure}
    \subfigure[$ $]{\hspace{-0.7cm}\includegraphics[scale=0.33]{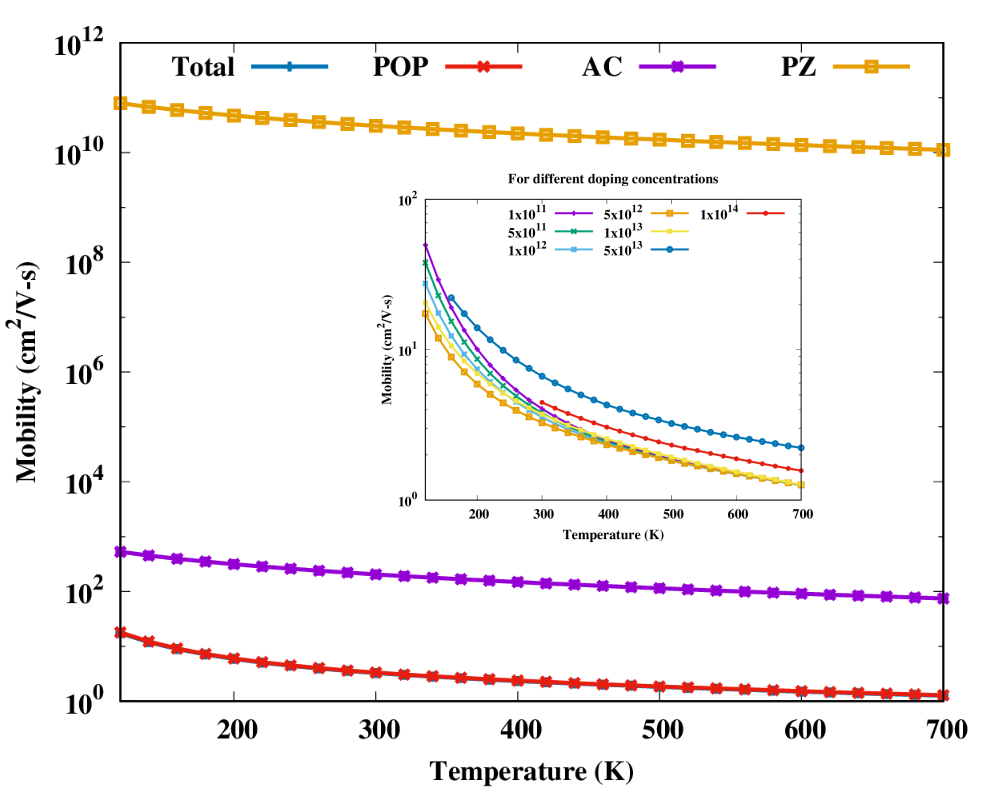}} 
    \subfigure[$ $]{\includegraphics[scale=0.32]{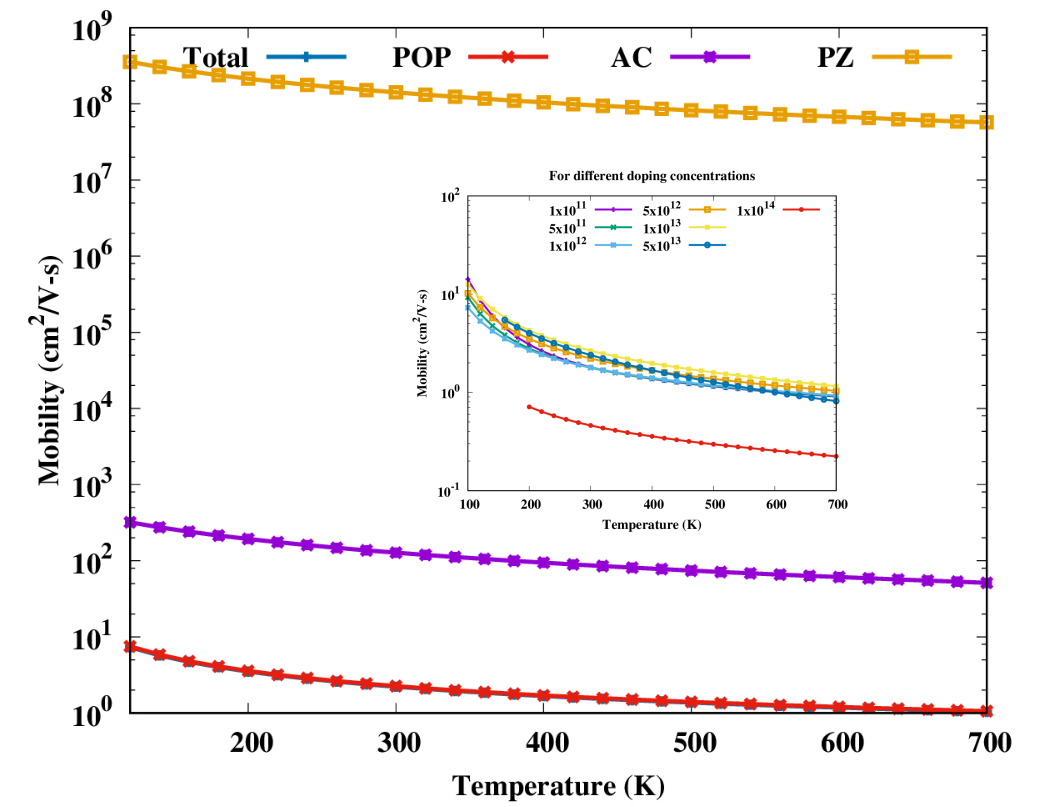}}\\
    \subfigure[$ $]{\hspace{-0.2cm}\includegraphics[scale=0.33]{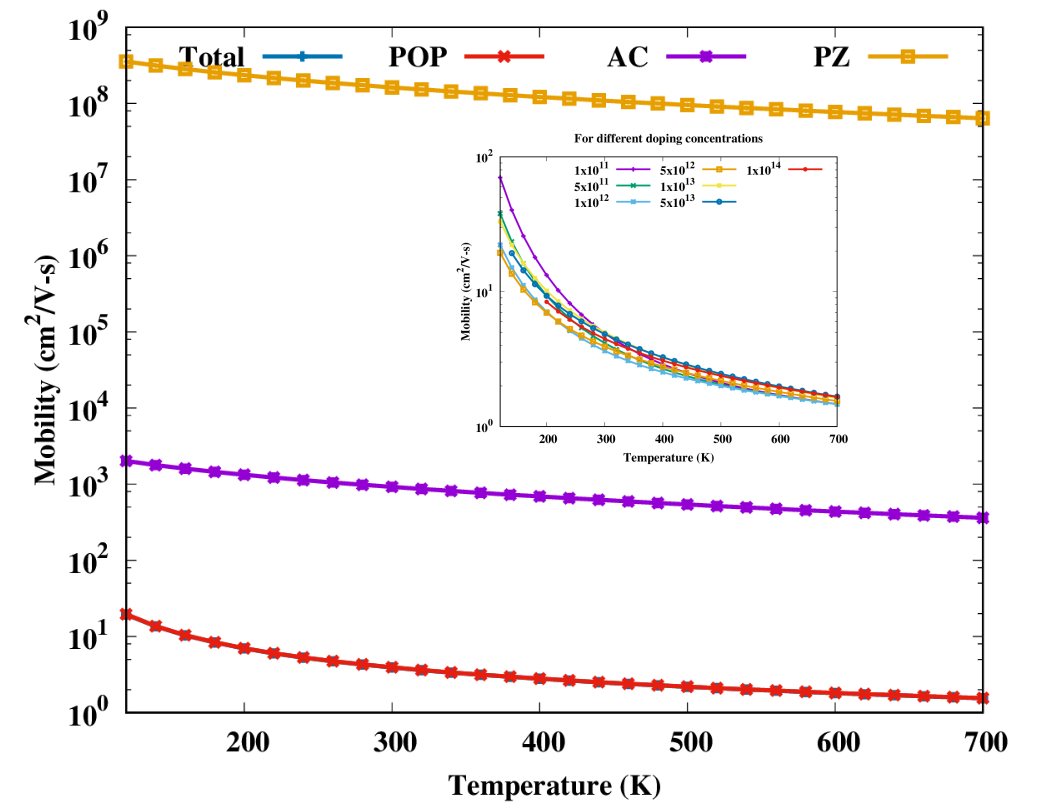}}
    \caption{Mobility of (a) \ch{Sc2CF2}, (b) \ch{Sc2CO2} and (c) \ch{Sc2C(OH)2} as a function of temperature for an electron concentration of 5$\times$10$^{12}$ cm$^{-2}$ with the contributions of POP, acoustic and PZ scattering. Inset shows the variation of total mobility with temperature at different electron concentrations in cm$^{-2}$.}
    \label{mobility}
\end{figure} 

The variation of conductivity as a function of temperature at different carrier concentrations are given in Figure S4. The conductivity increases with increase in electron concentration. At n = 1$\times$10$^{14}$ cm$^{-2}$, there is decrease in conductivity of \ch{Sc2CO2}. The conductivity value exceeds 1000 S/cm at high electron concentrations for \ch{Sc2CF2} and \ch{Sc2C(OH)2}. \\
\begin{figure}
    \subfigure[]{\hspace{-1.2cm}\includegraphics[scale=0.25]{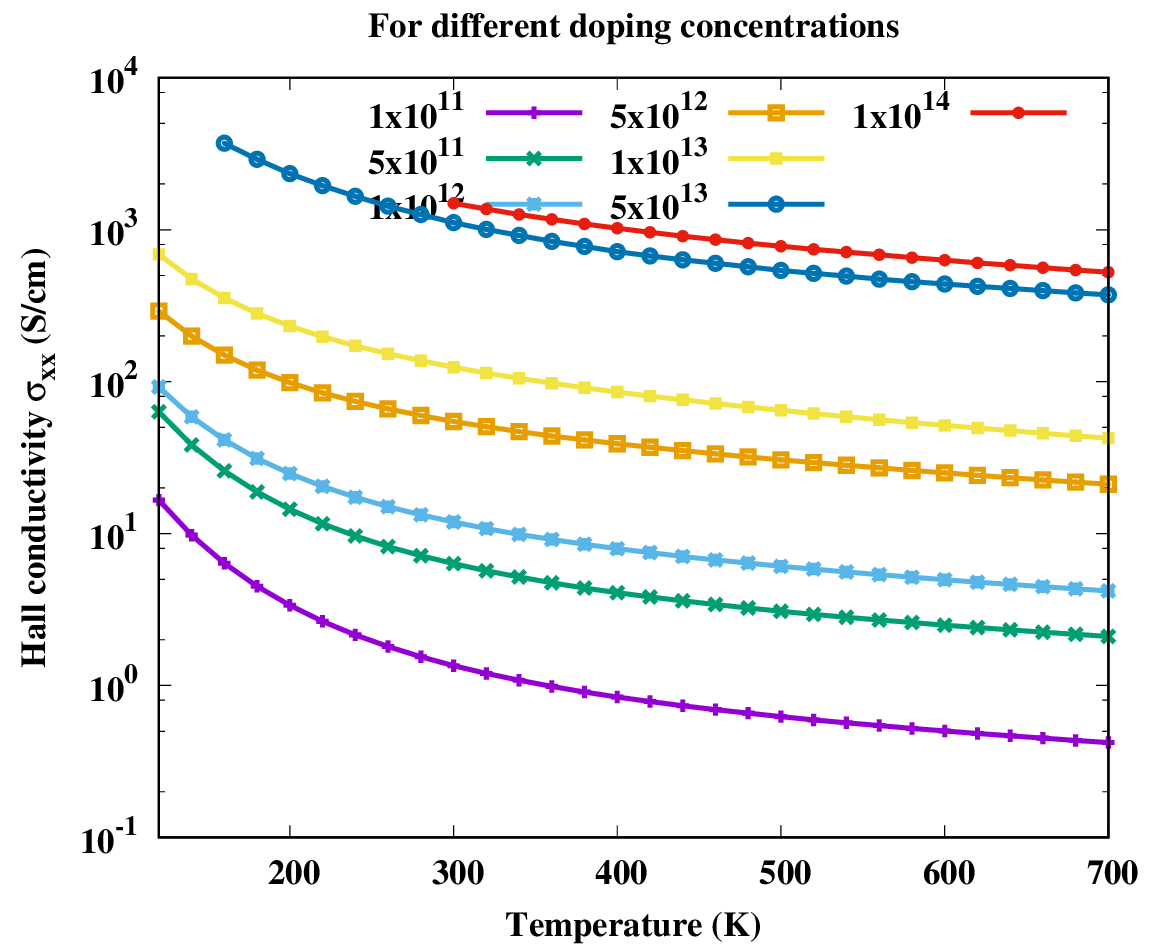}}
    \subfigure[]{\hspace{-0.2cm}\includegraphics[trim=18mm 0mm 0mm 0mm,clip,scale=0.25]{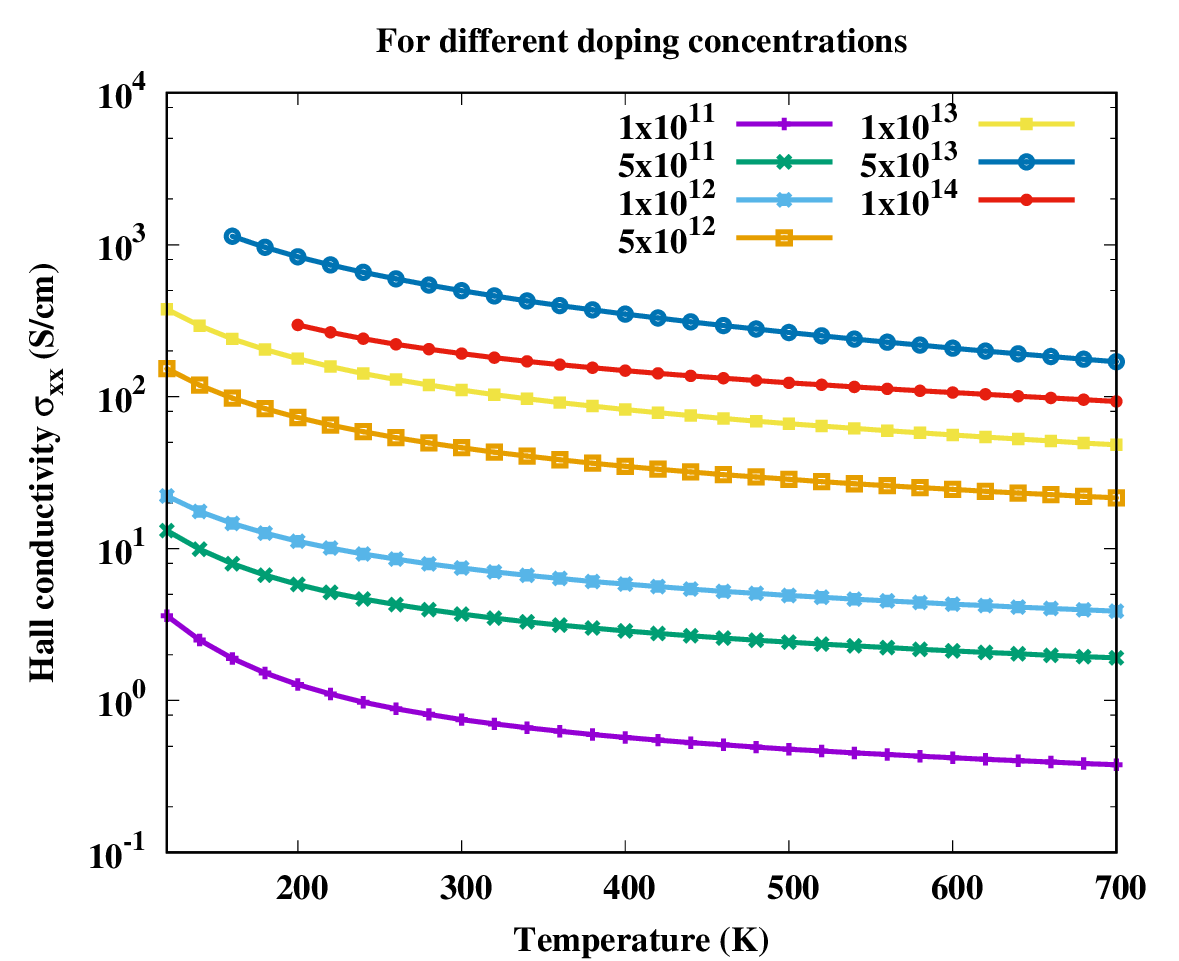}}
    \subfigure[]{\hspace{-0.2cm}\includegraphics[trim=18mm 0mm 0mm 0mm,clip,scale=0.25]{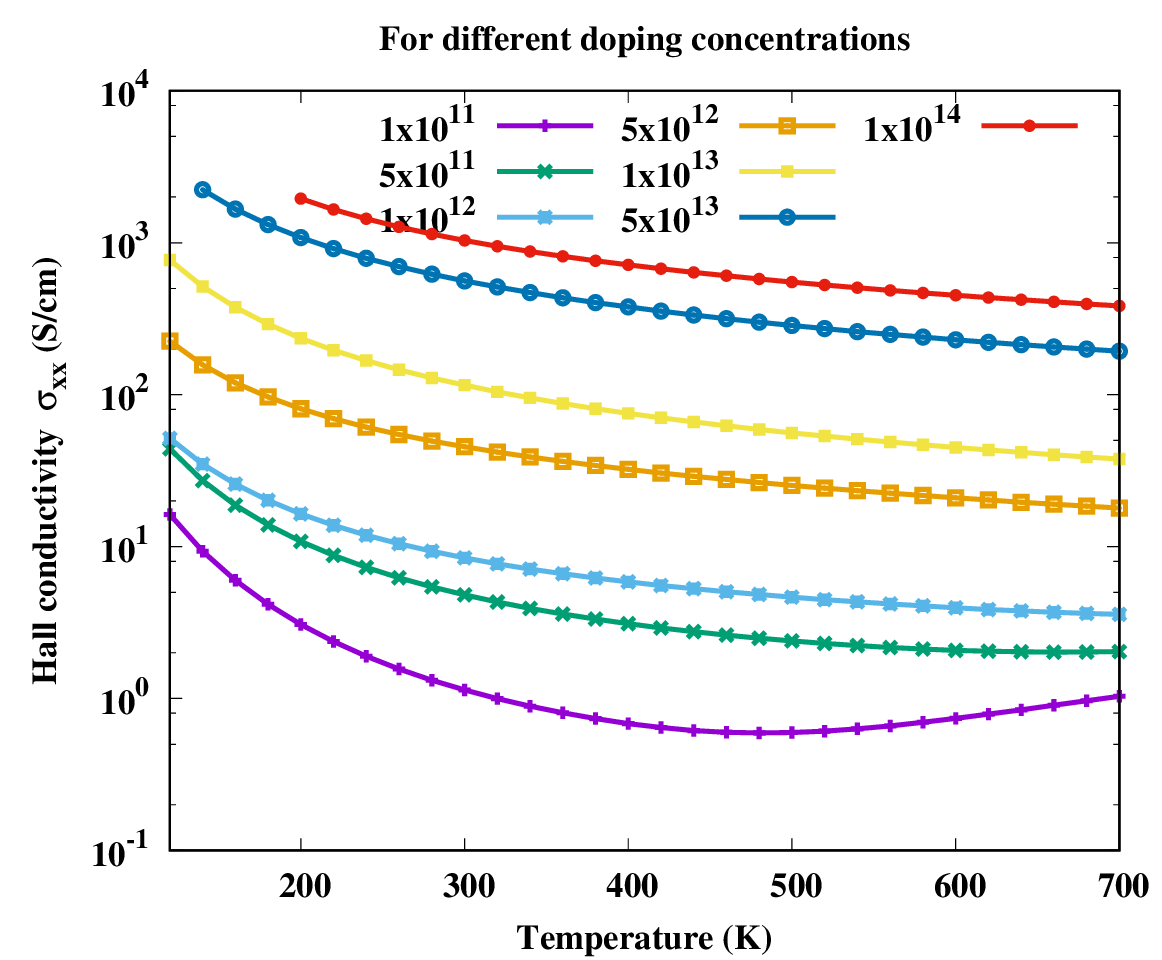}} \\
    \subfigure[]{\hspace{-1.2cm}\includegraphics[scale=0.25]{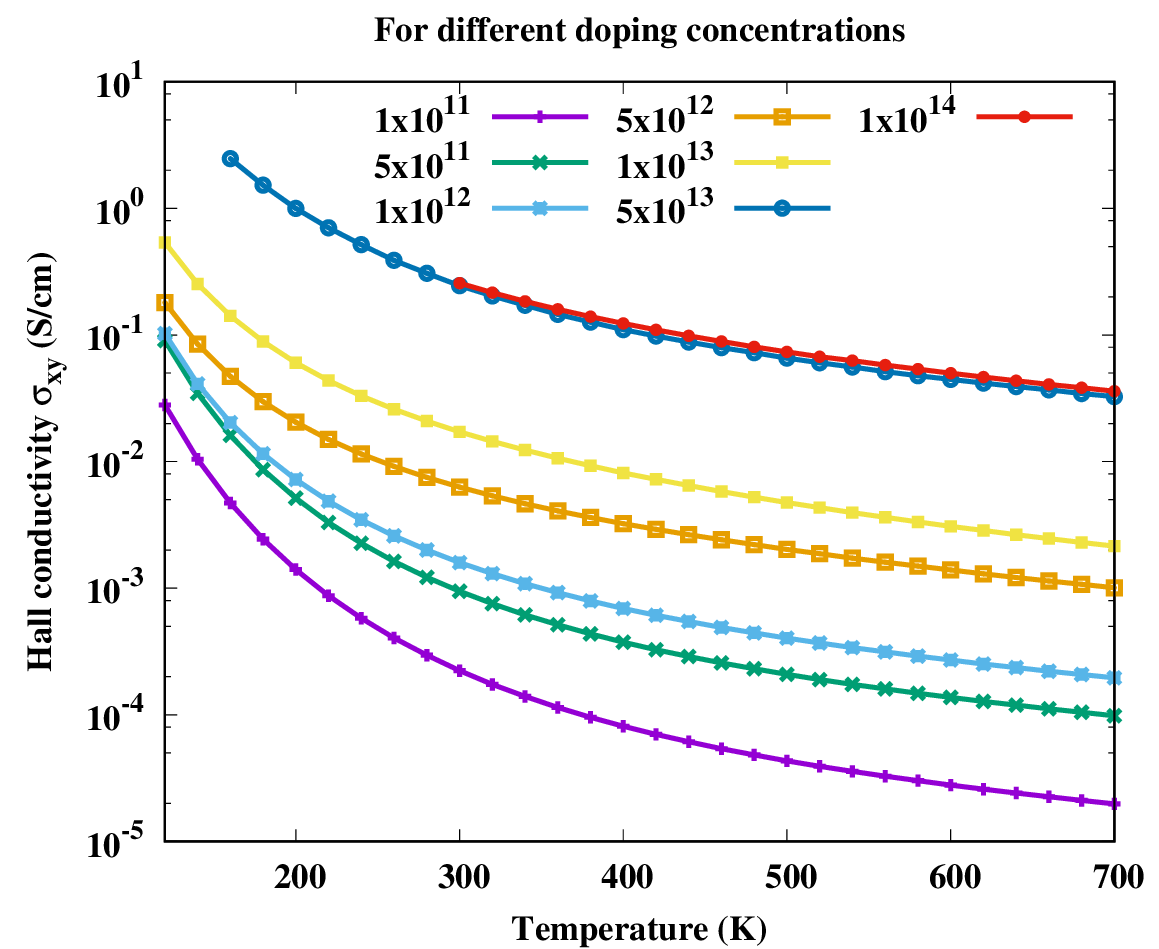}}
    \subfigure[]{\hspace{-0.15cm}\includegraphics[trim=14mm 0mm 0mm 0mm,clip,scale=0.25]{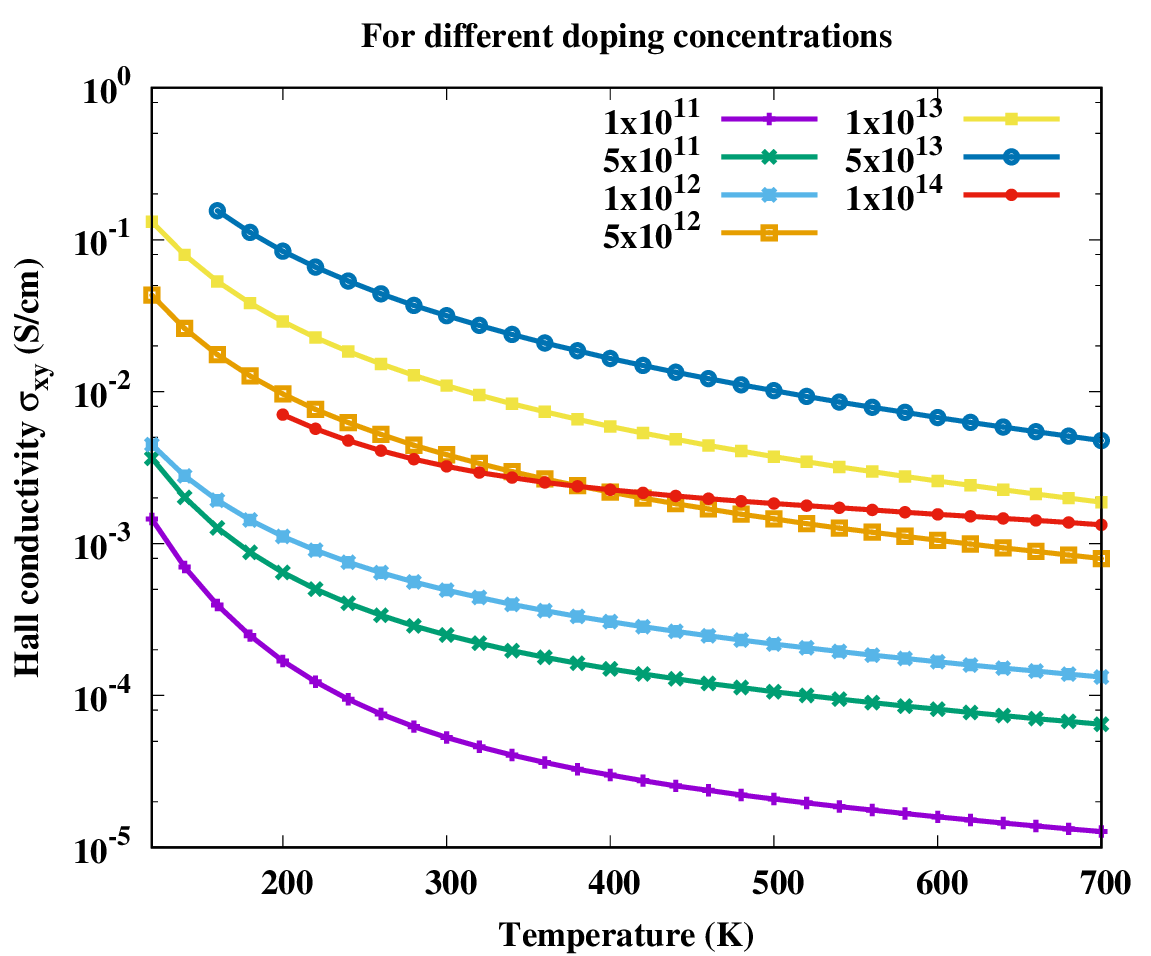}}
    \subfigure[]{\hspace{-0.15cm}\includegraphics[trim=14mm 0mm 0mm 0mm,clip,scale=0.25]{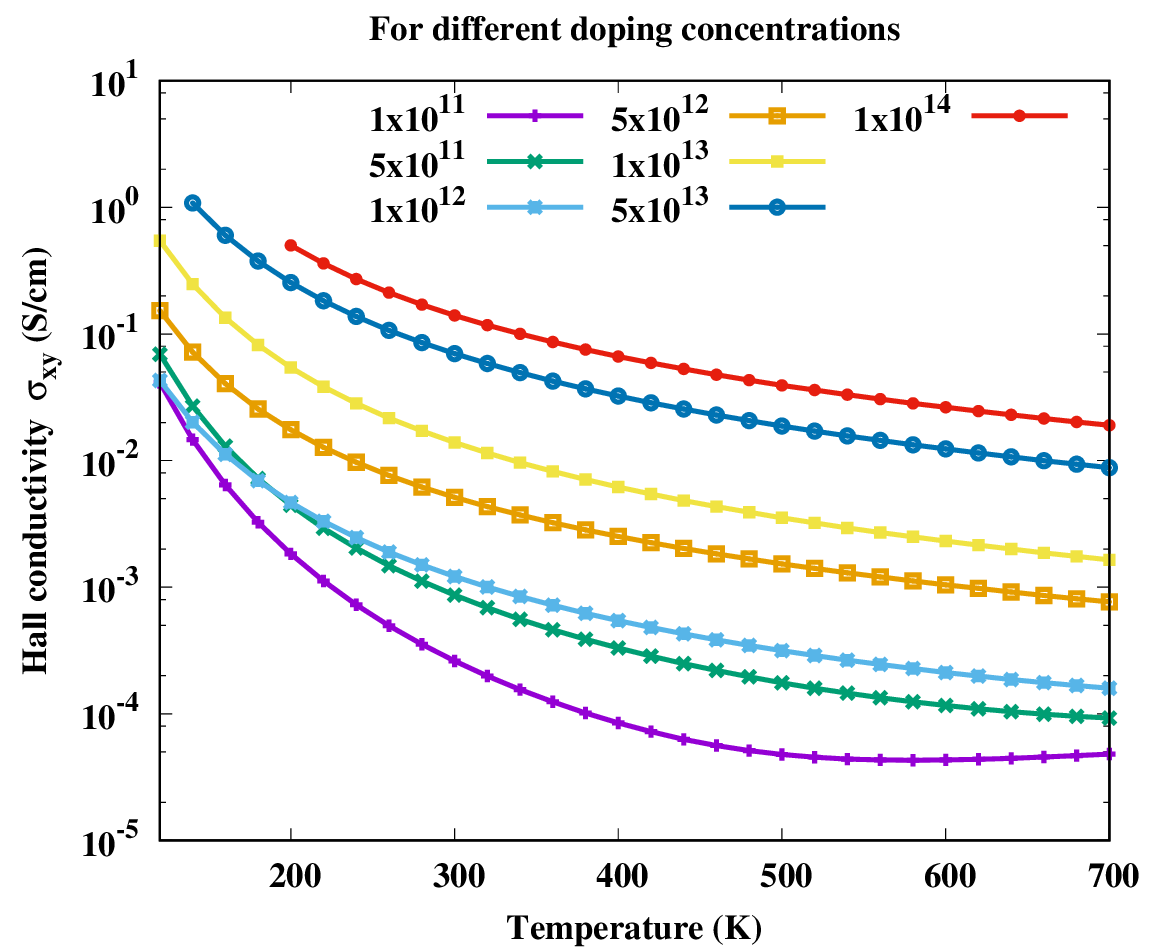}}
    \caption{Variation of Hall conductivity ($\sigma_{xx}$, $\sigma_{xy}$) of (a,d) \ch{Sc2CF2}, (b,e) \ch{Sc2CO2}, and (c,f) \ch{Sc2C(OH)2} with temperature at different electron concentrations. Here the first row corresponds to $\sigma_{xx}$ and the second row to $\sigma_{xy}$.}
    \label{Hall-cond}
\end{figure}

All magneto-transport calculations are performed at a small magnetic field of 0.4T and the field is along z-direction. The change in Hall conductivity ($\sigma_{xx}$ and $\sigma_{xy}$) of \ch{Sc2CF2}, \ch{Sc2CO2} and \ch{Sc2C(OH)2} as a function of temperature at different electron concentrations is shown in Figure \ref{Hall-cond}. As the carrier concentration increases, the Hall conductivity increases and it has the largest value at the highest electron concentration considered. For \ch{Sc2CO2}, the Hall conductivity decreases beyond electron concentration of 5$\times$10$^{13}$ cm$^{-2}$.  \\

\begin{figure}
    \subfigure[]{\hspace{-0.6cm}\includegraphics[scale=0.35]{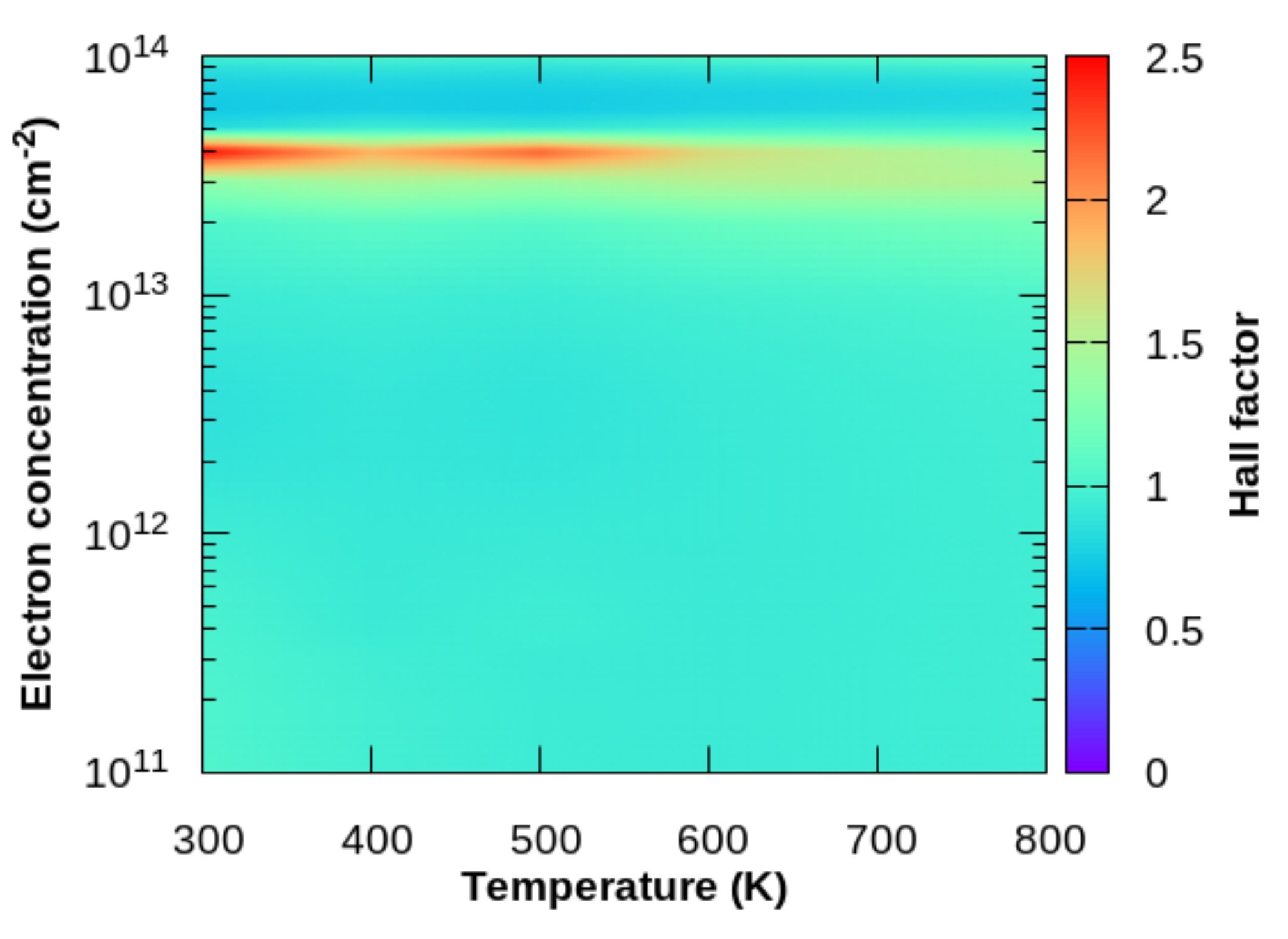}}
    \subfigure[]{\hspace{0cm}\includegraphics[scale=0.35]{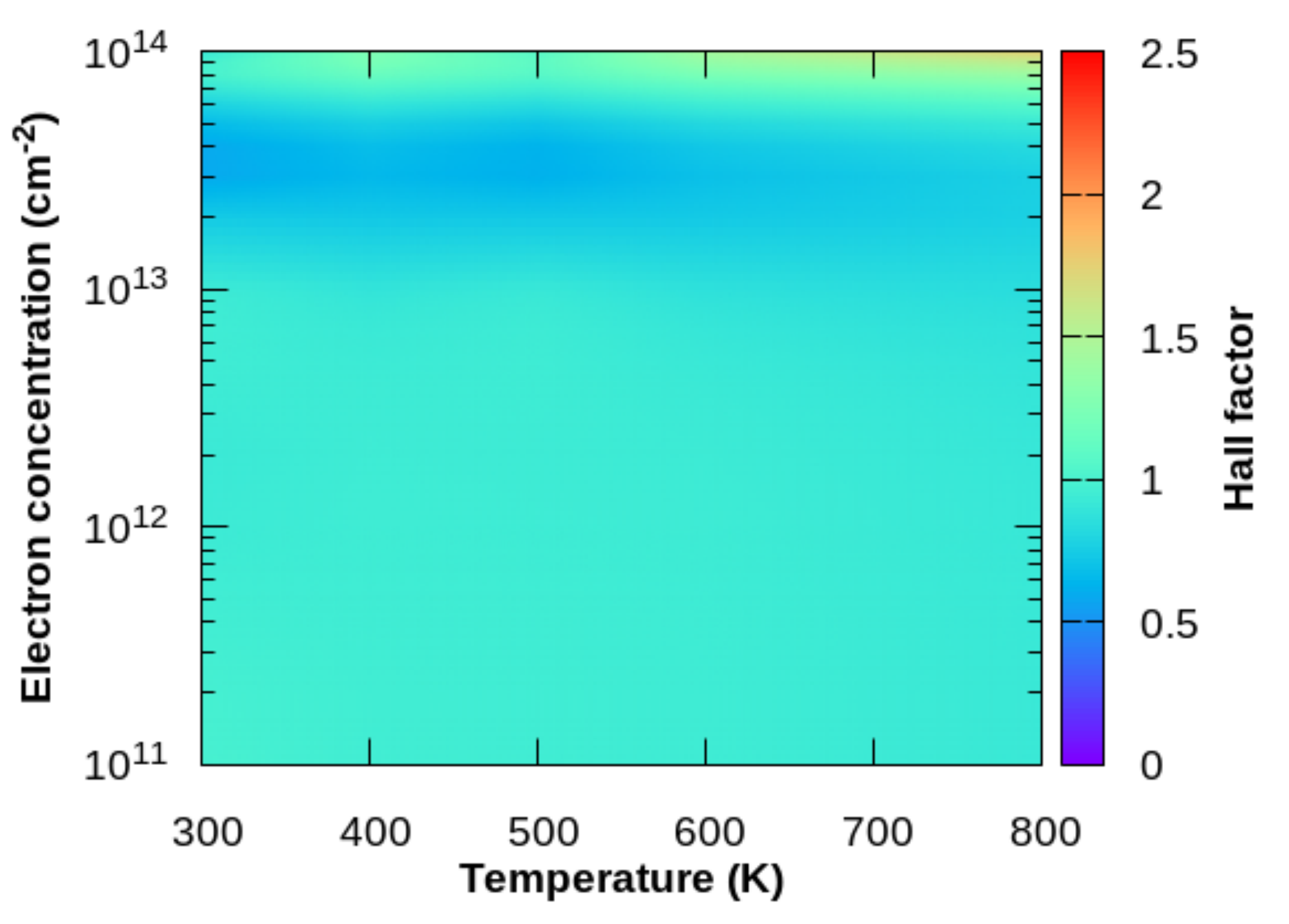}}
    \subfigure[]{\includegraphics[scale=0.35]{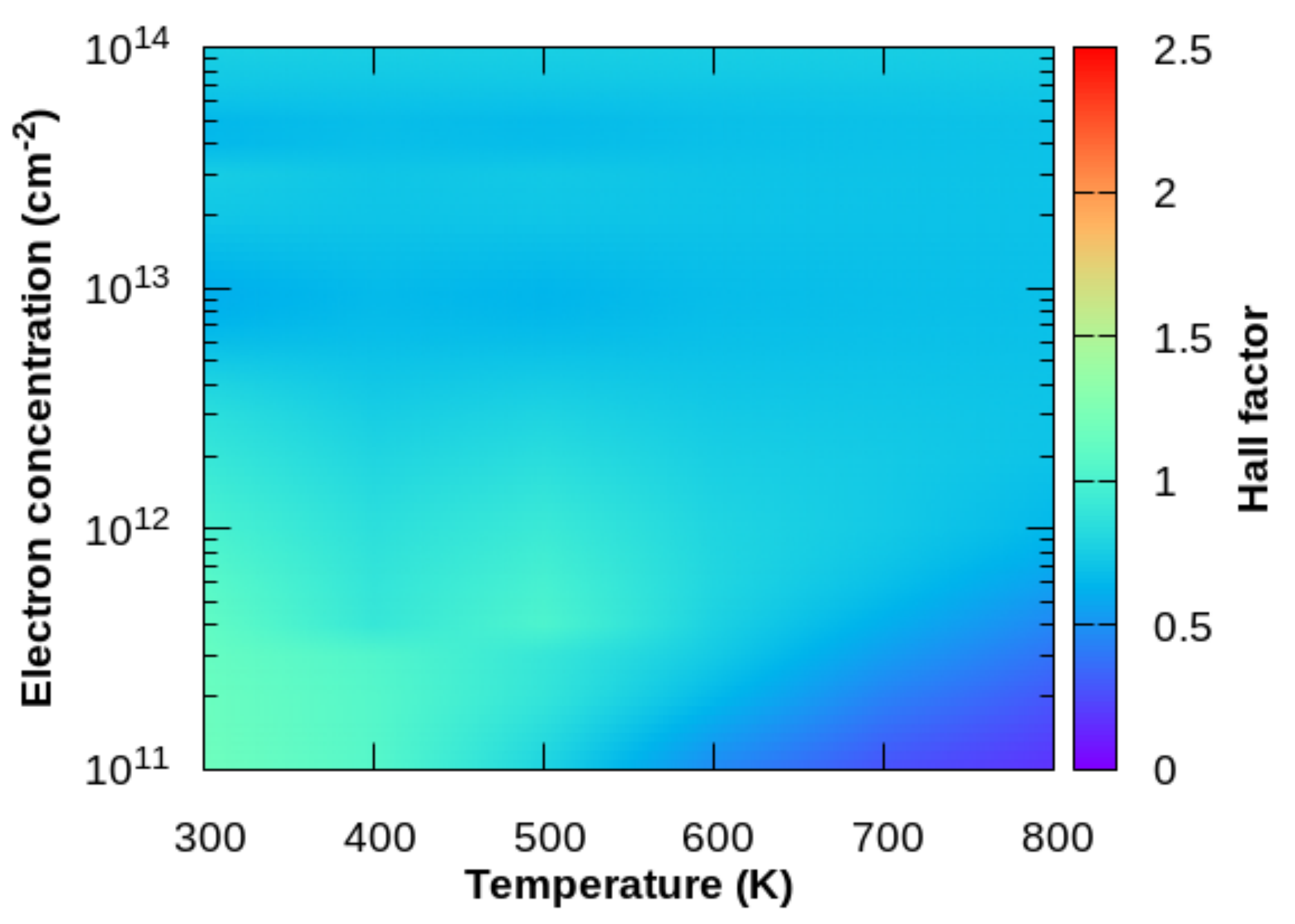}}
    \caption{Variation of the Hall factor in terms of temperature and concentration for (a)\ch{Sc2CF2}, (b)\ch{Sc2CO2} and (c)\ch{Sc2C(OH)2}.}
    \label{hall_final}
\end{figure}

In the Figure \ref{hall_final}, we show the calculated Hall factor for all the three compounds, \ch{Sc2CF2}, \ch{Sc2CO2} and \ch{Sc2C(OH)2} both as a function of temperature and electron concentration. It is important to note here that Hall factor formulated by us~\cite{mandia2022electrical} as shown in Eq.\ref{r} depends on both the perturbations $g(\varepsilon)$ as well as $h(\varepsilon)$. Such a method is quite distinct in comparison to the similar other currently proposed methods such as for example adopted by Macheda \textit{et al.}~\cite{macheda2018magnetotransport}, where the authors have calculated the Hall scattering factor in graphene in terms of the solutions of the Boltzmann transport equation. However, they used relaxation time approximation (RTA) and the effect of magnetic field on the distribution function was not considered in an explicit way as we have done. At a carrier concentration of 5$\times$10$^{12}$ cm$^{-2}$ and magnetic field of 0.4 T along the z-direction, the difference in Hall factor obtained using RTA and Rode's iterative method ($\Delta$r = r$_{RTA}$ - r$_{Rode}$) is presented in supporting information (Figure S5). We see that, at low temperatures, the difference $\Delta$r is large and it reduces around room temperature. Beyond room temperature, $\Delta$r increases slightly for \ch{Sc2C(OH)2} whereas it further reduces for \ch{Sc2CO2}.\\

For the three MXenes considered, the Hall scattering factor at different temperatures crossover with increasing carrier concentrations (which is more evident in xy plots in supporting information - Figure S6). The Hall factor varies over a large range with carrier concentration for \ch{Sc2CF2} when compared to \ch{Sc2CO2} and \ch{Sc2C(OH)2}. At a carrier concentration of 4$\times$10$^{13}$ cm$^{-2}$, \ch{Sc2CF2} has the highest value of r, which is close to 2.50. At low doping (carrier density), the Hall factor is around 1 for both \ch{Sc2CF2} and \ch{Sc2CO2}. For F functional group, it increases sharply with increasing electron concentration at lower temperatures. In the case of O functional group, r values dip and at the highest electron concentration considered (1$\times$10$^{14}$ cm$^{-2}$),  it ranges from 0.9 to 1.6. For \ch{Sc2C(OH)2}  at small doping and low temperature, r has a value of 1.17 and it drops to 0.76 with increasing temperature.  \\

At low doping concentrations, we see that Hall factor of \ch{Sc2C(OH)2} is dependent on temperature and decreases with increasing temperature, though the bands at the extrema have quasi-parabolic nature. Also, the r value increases with temperature at high carrier concentration for \ch{Sc2CO2}. These are not in favour of the general assumption in Hall measurements where r is taken to be unity (based on the weak dependence of r on temperature and scattering mechanisms in bulk semiconductors with quasi-parabolic bands \cite{macheda2020theory} and with relaxation time approximation). But for \ch{Sc2CF2} and \ch{Sc2CO2}, the Hall factor ranges around unity with no significant temperature dependence at low carrier concentrations.  \\

For all the three MXenes, we observe an unique behaviour in the carrier concentration region between 0.3 and $0.4 \times 10^{14}$ cm$^{-2}$. It is found that \ch{Sc2CF2} has the highest r value in this region and it diminishes with increasing temperature. Similarly for \ch{Sc2C(OH)2}, there is a faint region of moderate r value which decreases with temperature in the specified carrier concentration range. The obtained r values are the lowest in that region for \ch{Sc2CO2}, which increases slightly with temperature. This similarity in F- and OH-functionalized \ch{Sc2C} could be attributed to the way in which they affect electronic structure as they receive only one electron from the MXene surface whereas O group obtains two electrons. \\

Such unique behaviour can further be understood in terms of Eq.\ref{r} and the Table S3 of the supplemental material. From the Eq.\ref{r}, it can be seen that the Hall factor depends on the ratio $\frac{h(E)}{(g(E))^2}$ of the two distribution functions. It can be seen from the Table S3 that in the region where the carrier concentration varies between 0.3 and $0.4 \times 10^{14}$ cm$^{-2}$,  this ratio is largest at the Fermi energy ($\frac{h(E_F)}{(g(E_F))^2}$) for  \ch{Sc2CF2} followed by \ch{Sc2C(OH)2} and \ch{Sc2CO2}. Since the Fermi energy is the most relevant energy for the transport properties, the value of this ratio at this energy gives us a lot of insight. Such a behaviour tells us that the relative size of the $h(E)$ in comparison to $g(E)$ is largest for \ch{Sc2CF2}. This also demonstrates the necessity of splitting of the total distribution function to an electronic and magnetic part This also shows the need to split the total distribution function into an electronic and a magnetic part when the carriers are subjected to both electric and magnetic fields. 

Furthermore, It can be seen that F- and OH-functionalized \ch{Sc2C} show weaker temperature dependence in comparison to O-functionalized one. The signature of Hall factor as a function of carrier concentration and temperature for the three considered scandium carbide MXenes are therefore different and unique, which makes it a promising tool to distinguish the samples with varying concentrations of these three surface functional groups. The above arguments should be applicable for other semiconducting MXenes as well, and Hall factor measurement may provide a simple solution to the long-standing challenge of determining the surface termination groups experimentally.

%
%
\newpage
\bibliography{Ref}
\subsubsection{}


\section{Acknowledements}
 AKM and BM gratefully acknowledge funding from Indo-Korea Science and Technology Center (IKST), Bangalore.

\section{Conflict of interest}
The authors declare that they have no conflict of interest.
%

\end{document}


\begin{figure}
\subfigure[$ $]{\hspace{-0.5cm}\includegraphics[scale=0.5]{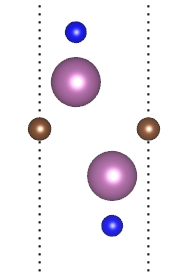}} 
\subfigure[$ $]{\hspace{0cm}\includegraphics[trim=0mm 0mm 0mm 5mm,clip,scale=0.5]{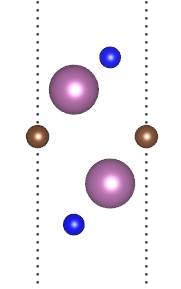}}
\subfigure[$ $]{\hspace{0cm}\includegraphics[scale=0.5]{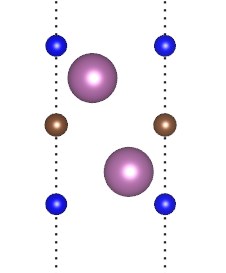}}
\subfigure[$ $]{\hspace{0cm}\includegraphics[trim=10mm 0mm 0mm 0mm,clip,scale=0.5]{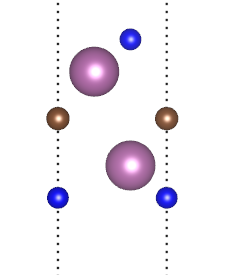}} \\
\subfigure[$ $]{\hspace{-1cm}\includegraphics[scale=0.5]{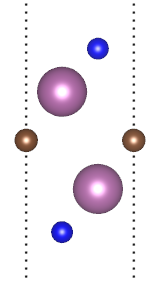}}
\subfigure[$ $]{\hspace{0.8cm}\includegraphics[scale=0.48]{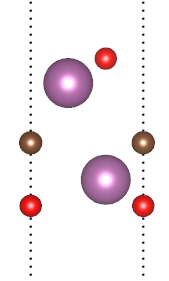}} 
\subfigure[$ $]{\hspace{0.8cm}\includegraphics[scale=0.52]{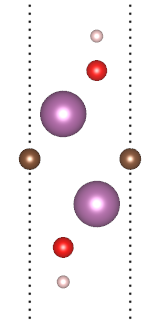}}\\
\subfigure[$ $]{\hspace{-0.5cm}\includegraphics[scale=0.52]{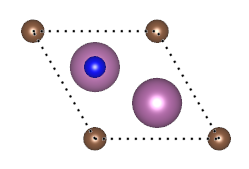}}
\subfigure[$ $]{\hspace{0cm}\includegraphics[scale=0.5]{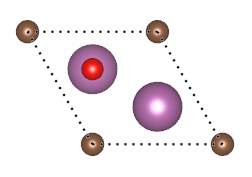}} 
\subfigure[$ $]{\hspace{0cm}\includegraphics[scale=0.55]{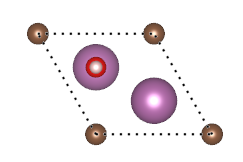}}\\
 \caption{Side view of \ch{Sc2CF2} (a) I, (b) II, (c) III and (d) IV configurations. Side and top view of optimized structure of \ch{Sc2C} functionalized by (e,h) F, (f,i) O and (g,j) OH. Lavendar, brown, blue, red and white balls correspond to Sc, C, F, O and H atoms respectively.}
 \label{figure_S1}
\end{figure}
%
\begin{figure}
    \subfigure[$ $]{\hspace{-0.7cm}\includegraphics[scale=0.65]{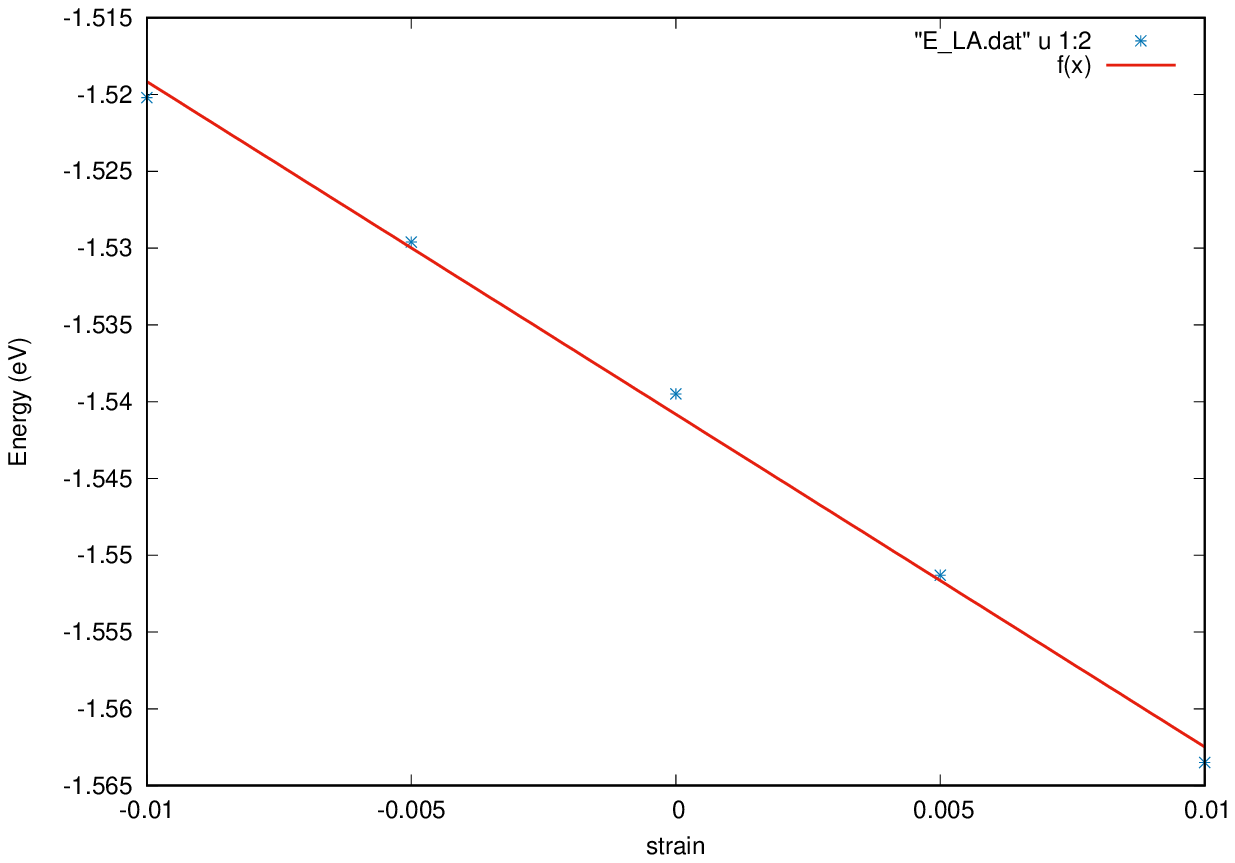}}
    \subfigure[$ $]{\includegraphics[scale=0.65]{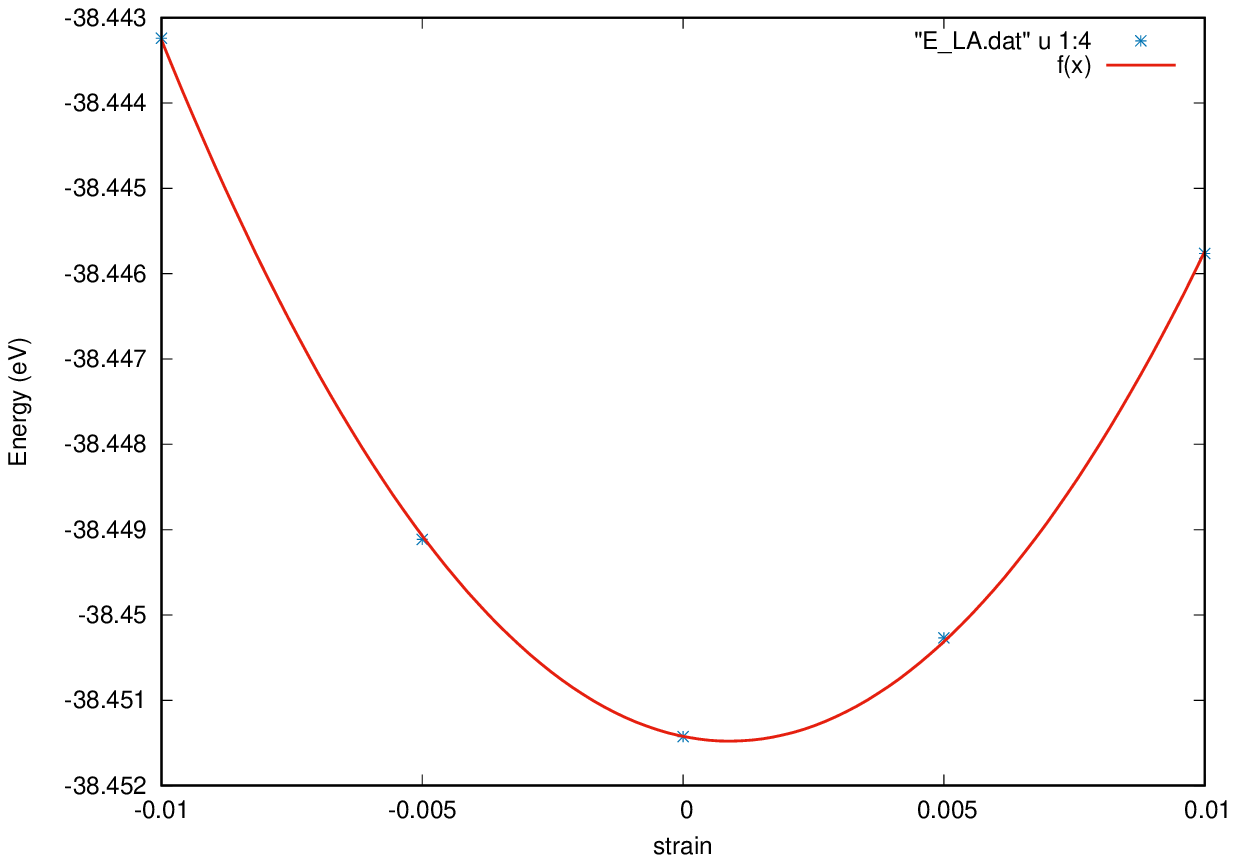}}\\
    \subfigure[$ $]{\hspace{-0.7cm}\includegraphics[scale=0.65]{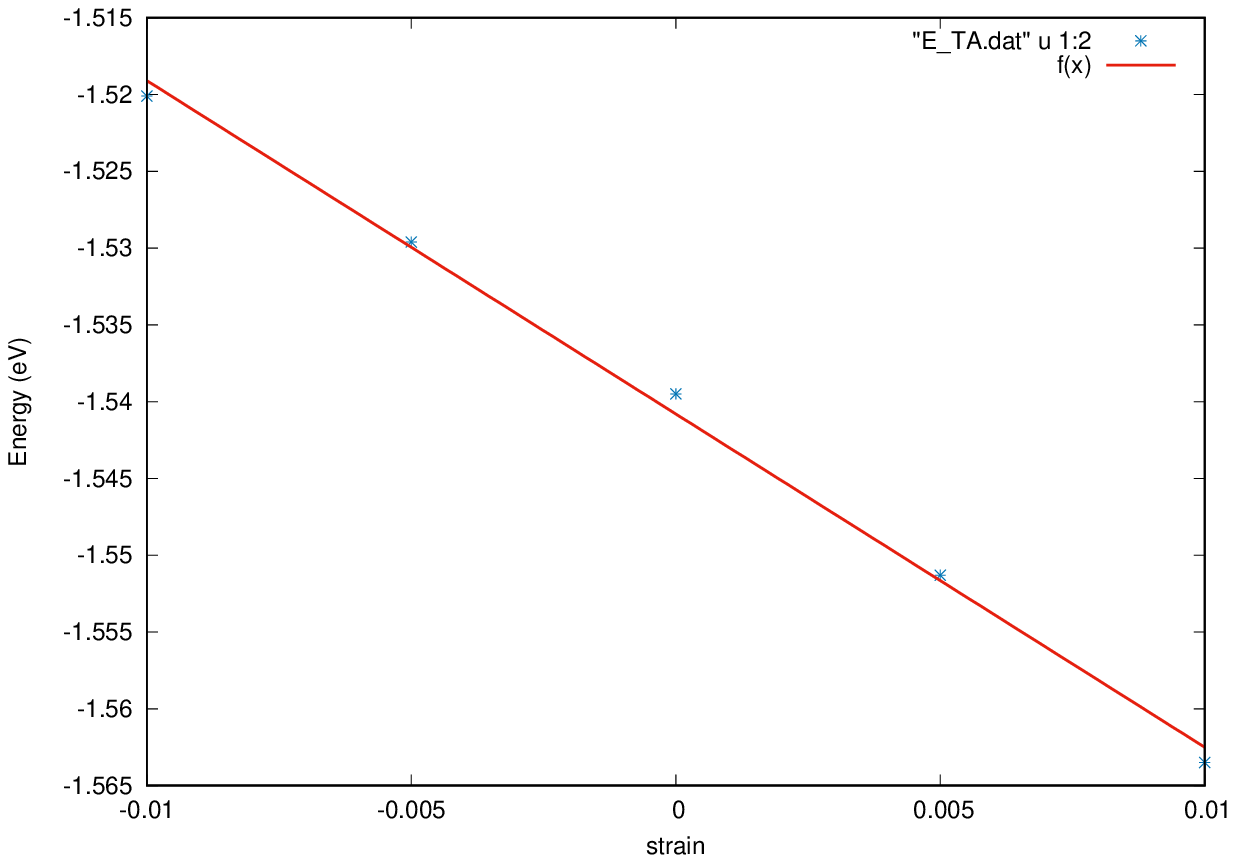}}
    \subfigure[$ $]{\includegraphics[scale=0.65]{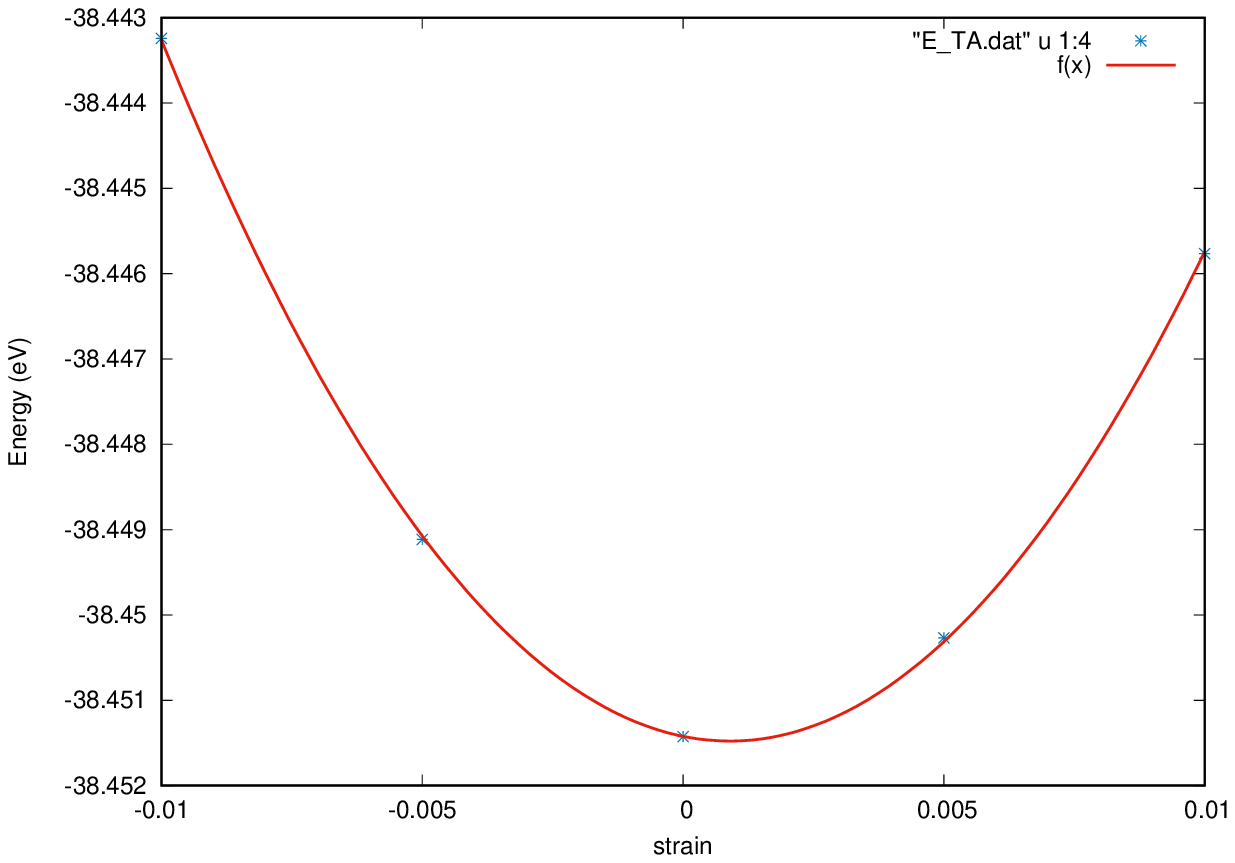}}\\
    \subfigure[$ $]{\hspace{-0.7cm}\includegraphics[scale=0.65]{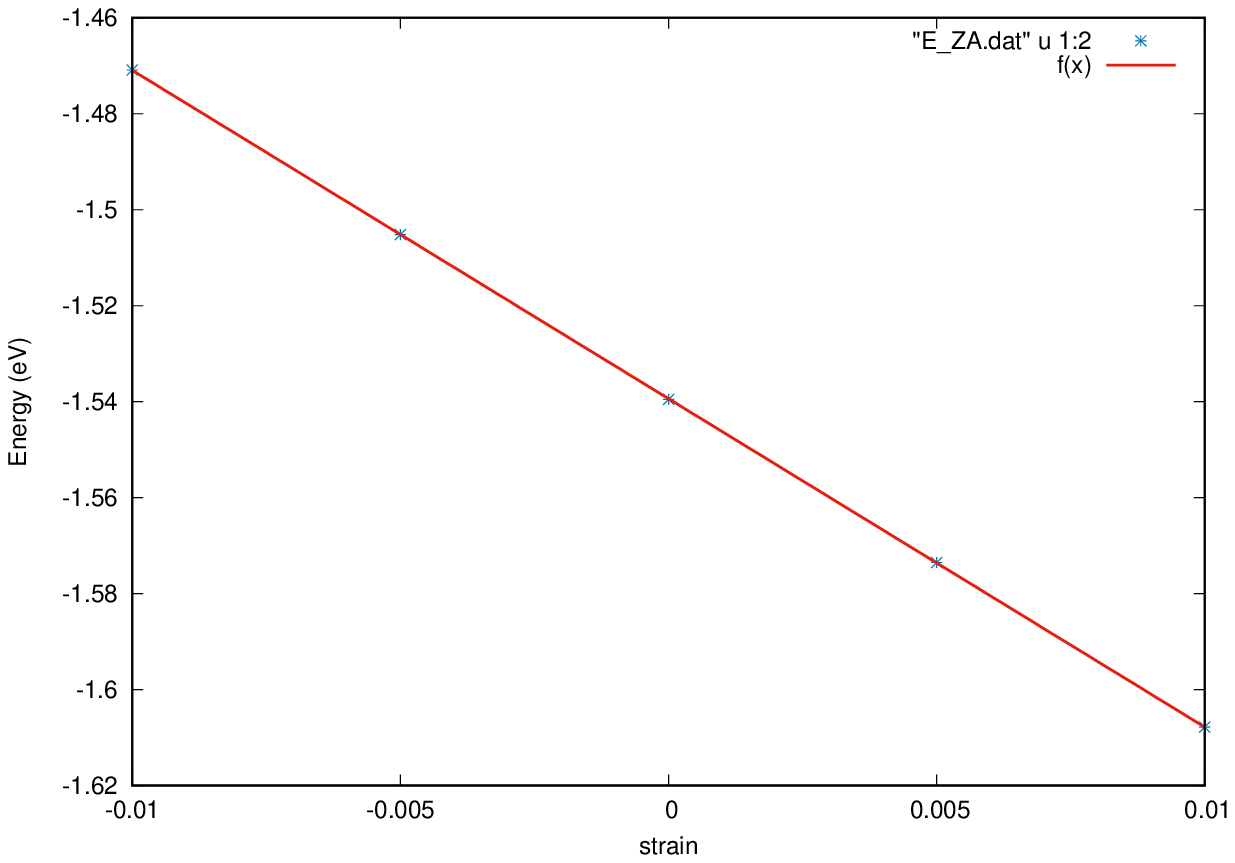}}
    \subfigure[$ $]{\includegraphics[scale=0.65]{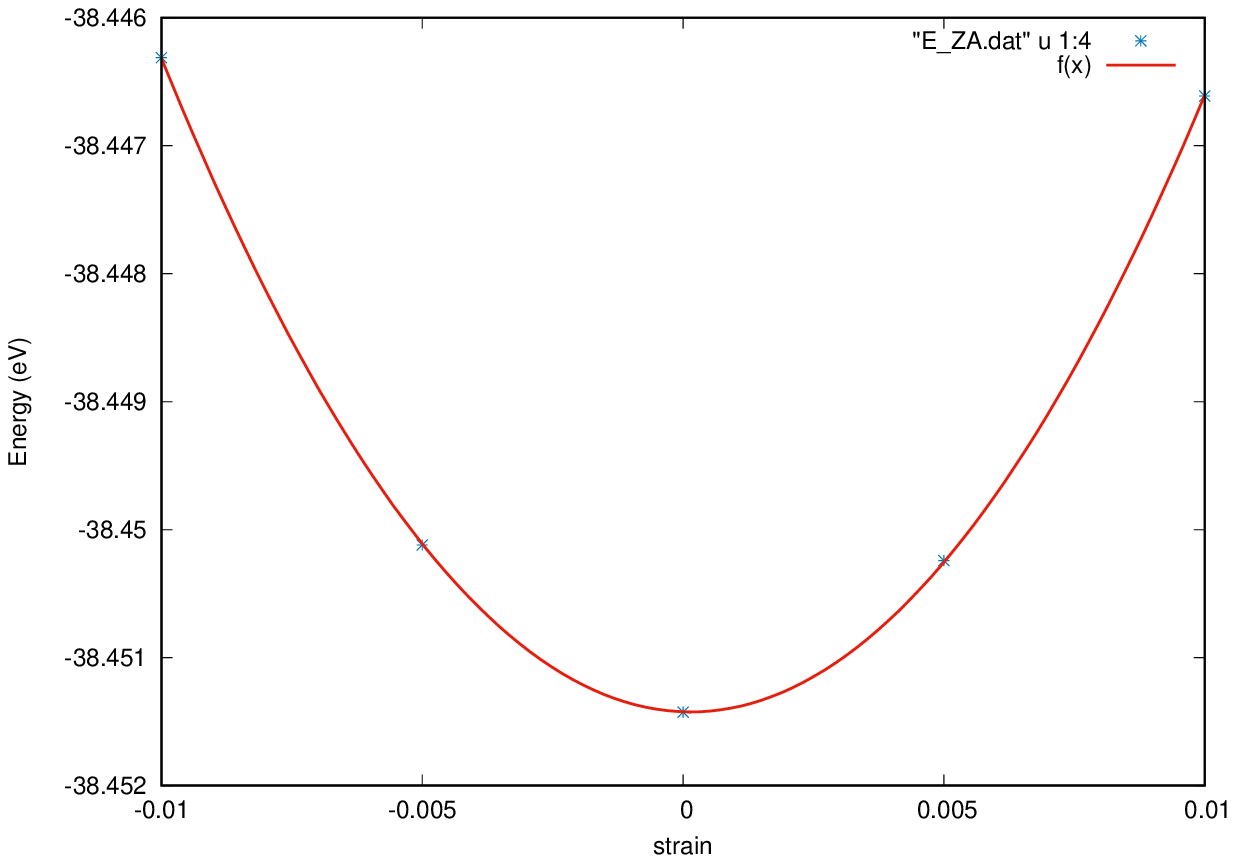}}
    \caption{For \ch{Sc2CF2}, the conduction band edge shift position for uniaxial strain along (a) x, (c) y, and (e) z directions. The relationship between total energy and strain along (b) x, (d) y, and (f) z directions are given for \ch{Sc2CF2}.}
    \label{CB_edge}
\end{figure}
\FloatBarrier
\begin{table}[ht]
\caption{Deformation potential calculation for \ch{Sc2CF2}: E($\epsilon$)=a$\epsilon$+b, D$_{A}$=$\frac{\partial E}{\partial \epsilon}$=a}
\begin{tabular}{ccccccccc}
& LA & TA  &  ZA   \\ 
\hline 
Parameter a &  -2.166 & -2.17 & -6.842 \\
Standard error & $\pm$0.07429 & $\pm$0.07243 & $\pm$0.0006 \\ \\
Parameter b & -1.541  & -1.541 &-1.539 \\ 
Standard error & $\pm$0.00053 & $\pm$0.00051 &$\pm$0.00004
\end{tabular}
\label{table1}
\end{table}
\FloatBarrier
\FloatBarrier
\begin{table}
\caption{Elastic moduli calculation for \ch{Sc2CF2}: E($\epsilon$)=a$\epsilon^{2}$+b$\epsilon$+c, C=$\frac{\partial^2 E}{\partial \epsilon^2}$=2a}
\begin{tabular}{ccccccccc}
& LA & TA  &  ZA   \\ 
\hline 
Parameter a &  69.218 & 69.205 & 49.620 \\
Standard error & $\pm$0.522 & $\pm$0.517 & $\pm$0.142 \\ \\
Parameter b & -0.124  & -0.124 & -0.014 \\ 
Standard error & $\pm$0.0031 & $\pm$0.0031 &$\pm$0.0008 \\ \\
Parameter c & -38.451  & -38.451 &-38.451 \\ 
Standard error & $\pm$0.00003 & $\pm$0.00003 &$\pm$0.000009
\end{tabular}
\label{table1}
\end{table}
\FloatBarrier
\begin{figure}[hbt!]
    \centering
    \subfigure[$ $]{\hspace{-1cm}\includegraphics[scale=0.25]{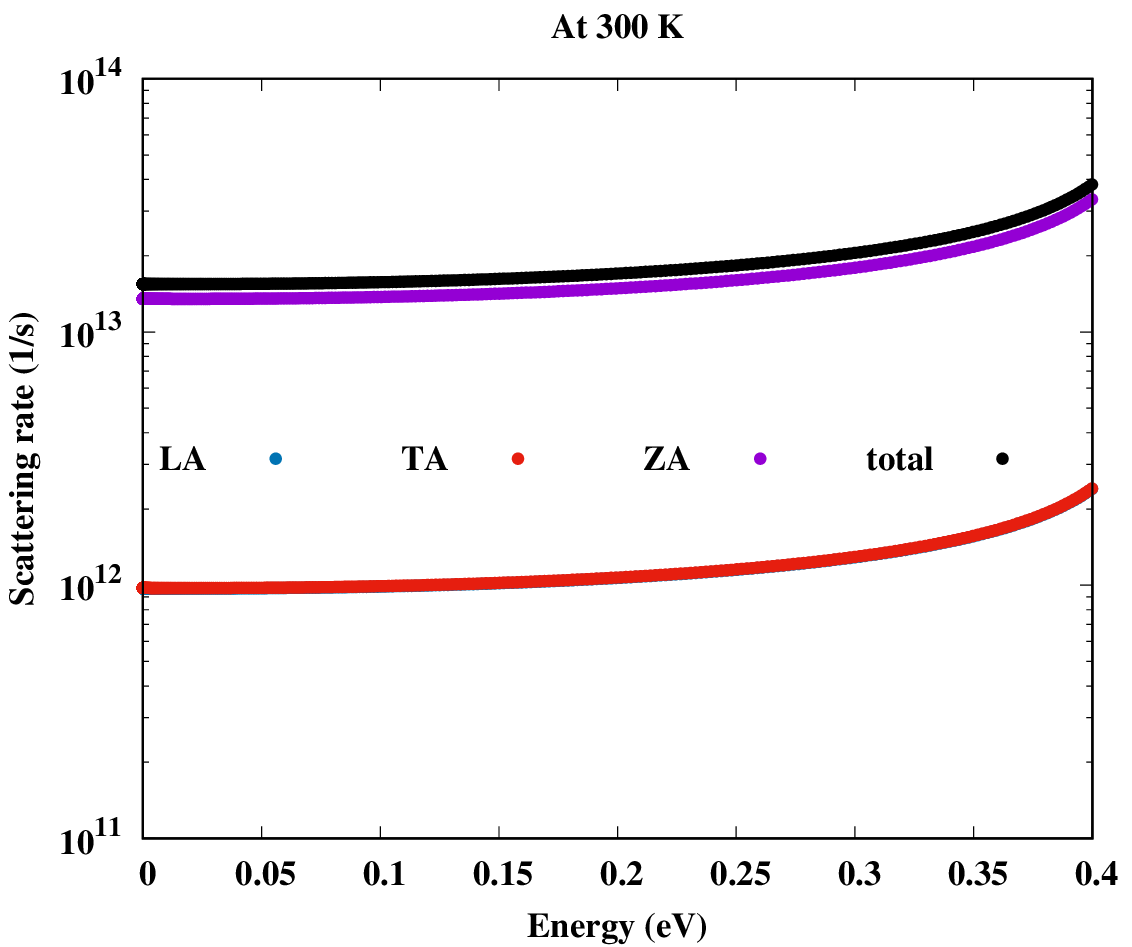}}
    \subfigure[$ $]{\hspace{-0.2cm}\includegraphics[trim=10mm 0mm 2mm 0mm,clip,scale=0.25]{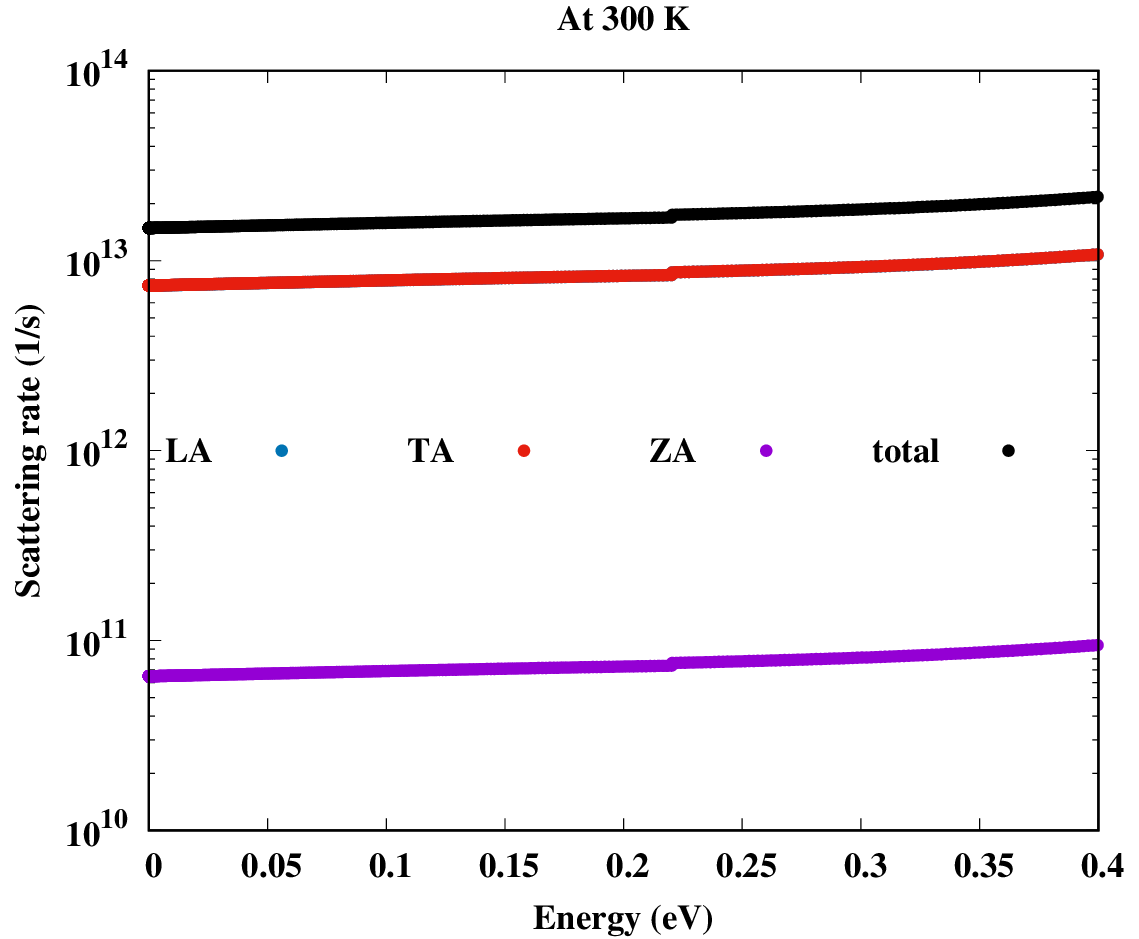}}
    \hfill
    \subfigure[$ $]{\hspace{-0.2cm}\includegraphics[trim=16mm 0mm 0mm 0mm,clip,scale=0.25]{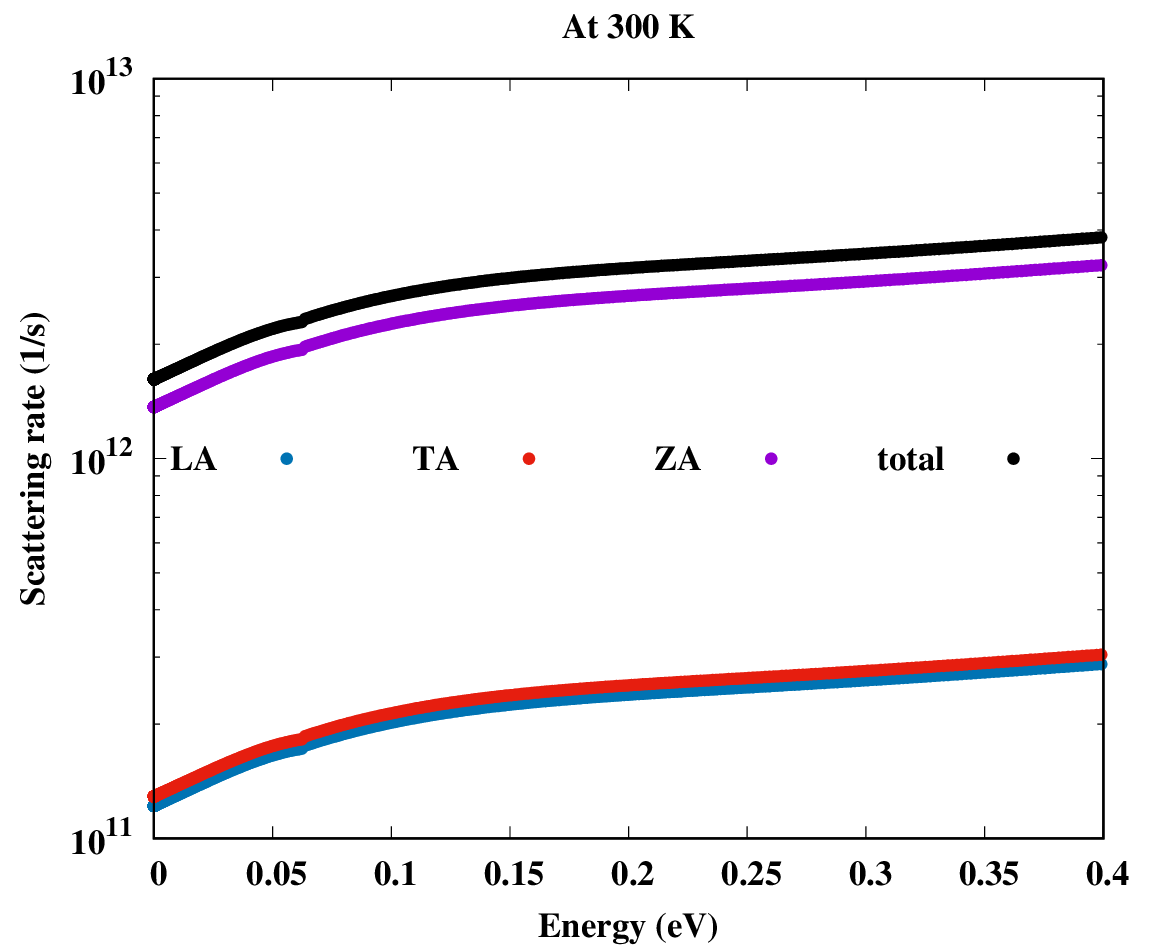}}
    \caption{Scattering rates versus energy due to acoustic phonons: (a) \ch{Sc2CF2}, (b) \ch{Sc2CO2} and (c) \ch{Sc2C(OH)2}.}
    \label{scattering_AC}
\end{figure}
\begin{figure}[hbt!]
    \centering
    \subfigure[$ $]{\hspace{-1.2cm}\includegraphics[scale=0.25]{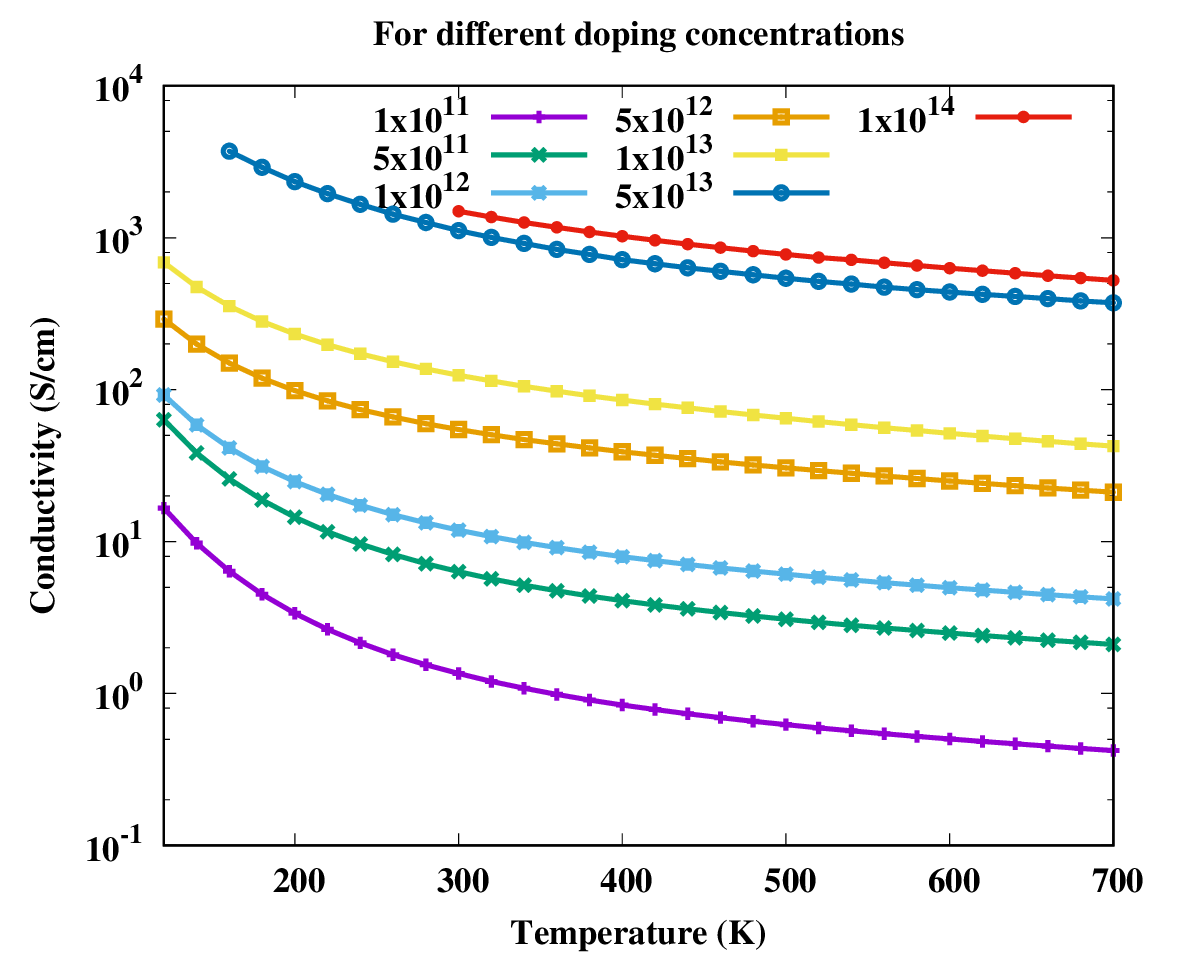}}
    \subfigure[$ $]{\hspace{-0.2cm}\includegraphics[trim=16mm 0mm 2mm 0mm,clip,scale=0.25]{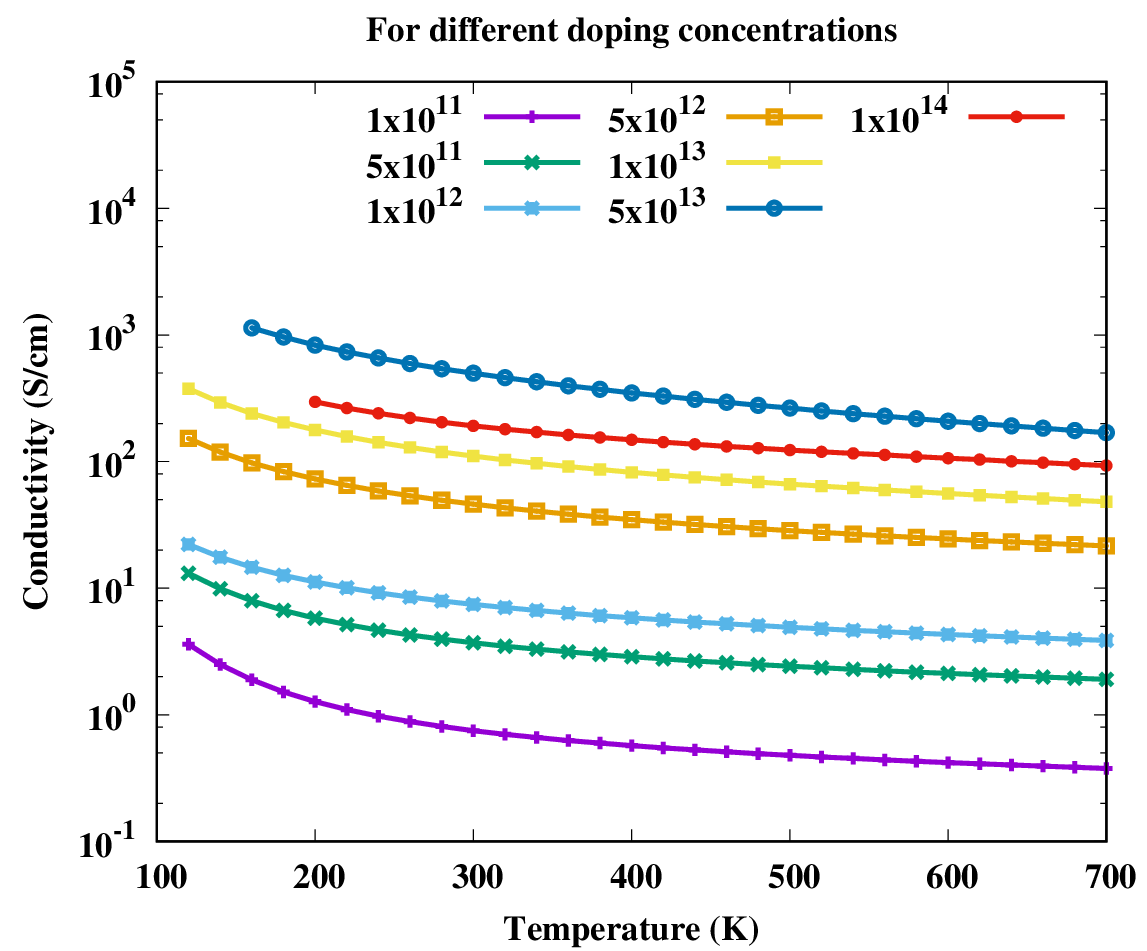}}
    \hfill
    \subfigure[$ $]{\hspace{-0.2cm}\includegraphics[trim=16mm 0mm 0mm 0mm,clip,scale=0.25]{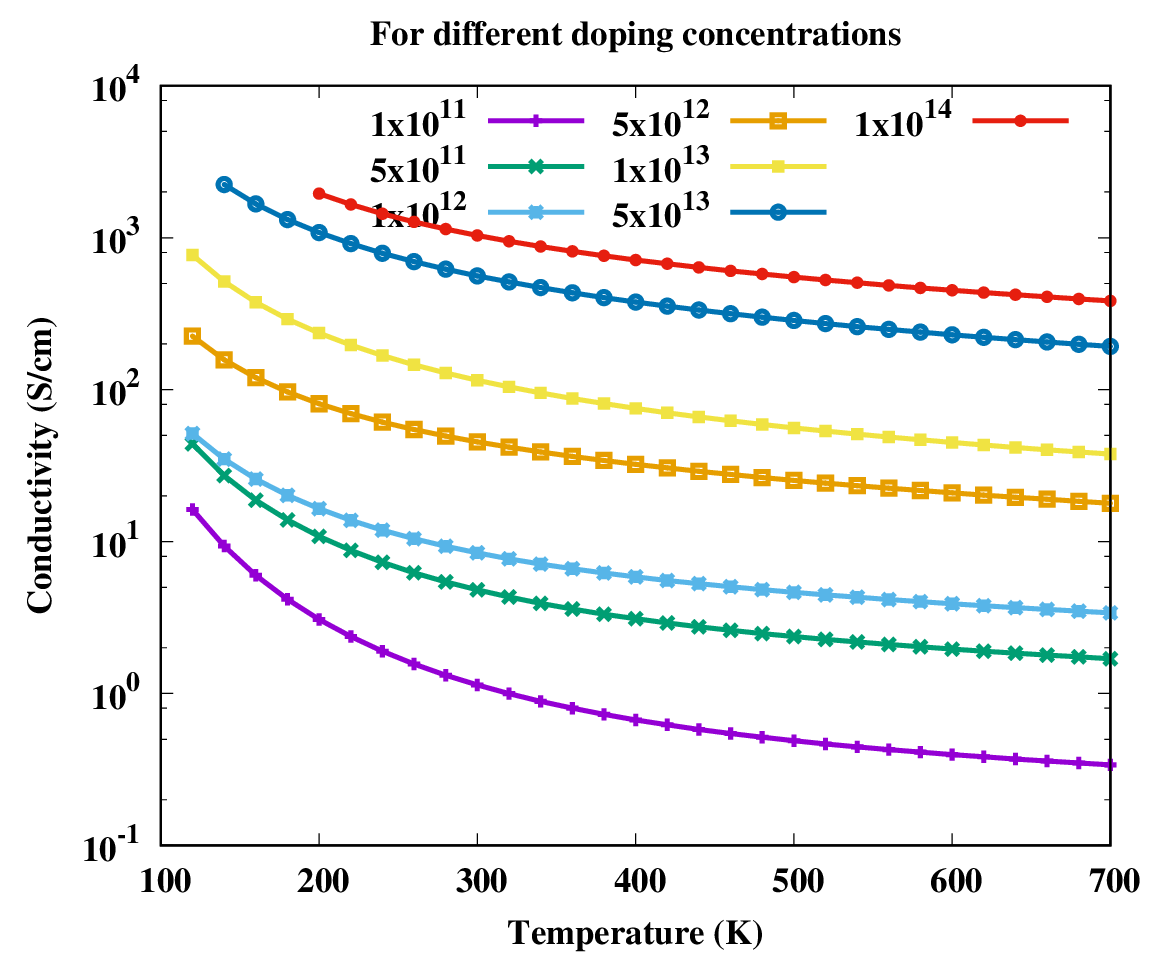}}
    \caption{Conductivity as a function of temperature: (a) \ch{Sc2CF2}, (b) \ch{Sc2CO2} and (c) \ch{Sc2C(OH)2}.}
    \label{scattering_AC}
\end{figure}
\begin{figure}
    \centering
    \subfigure[]{\includegraphics[scale=0.40]{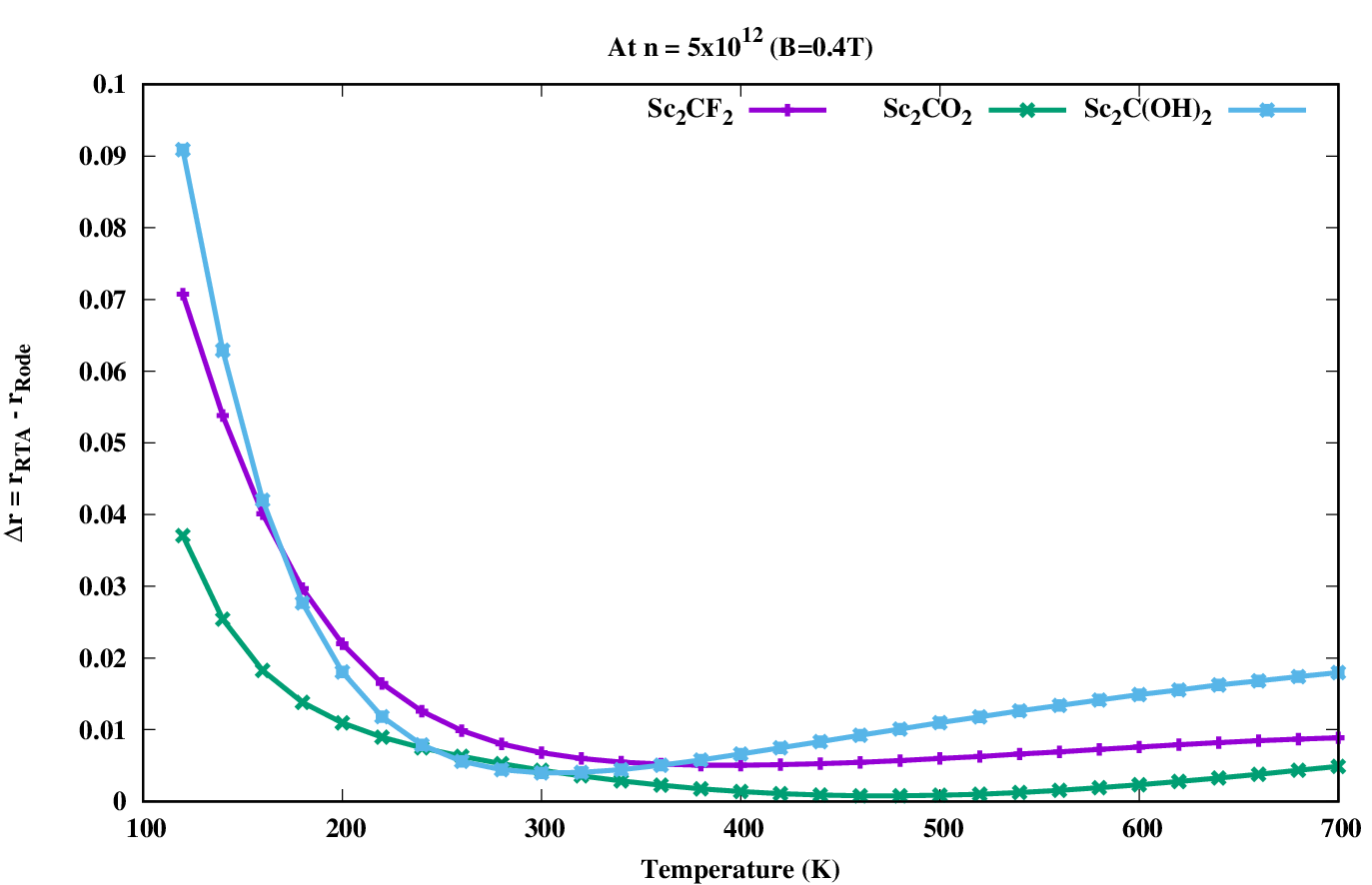}}
    \caption{For a given concentration (n=5$\times$10$^{12}$ cm$^{-2}$), the difference in Hall scattering factor $\Delta$r calculated using RTA and Rode approach as a function of temperature.}
    \label{Delta_r}
\end{figure}
\begin{figure}
    \centering
    \subfigure[$ $]{\hspace{-0.7cm}\includegraphics[scale=0.33]{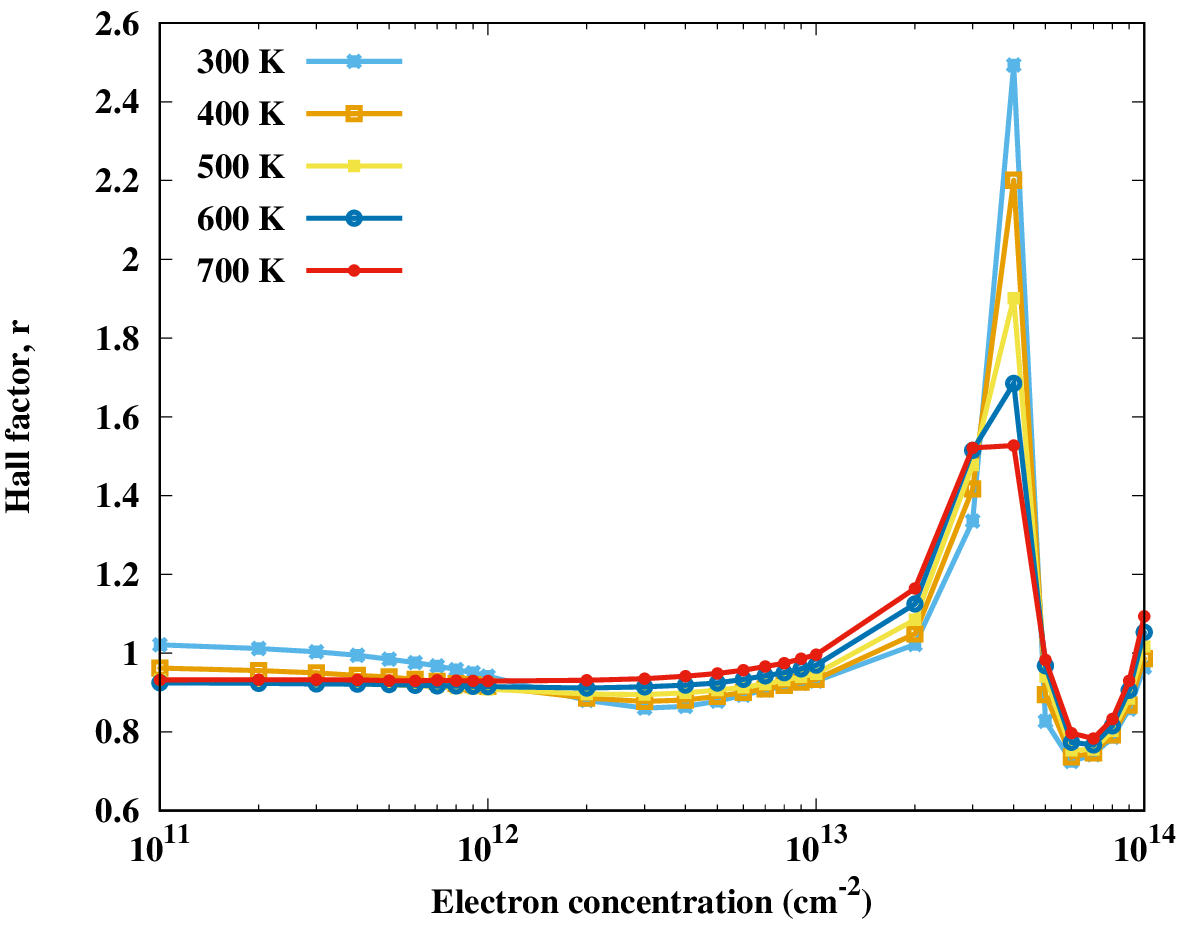}}
    \hfill
    \subfigure[$ $]{\includegraphics[scale=0.33]{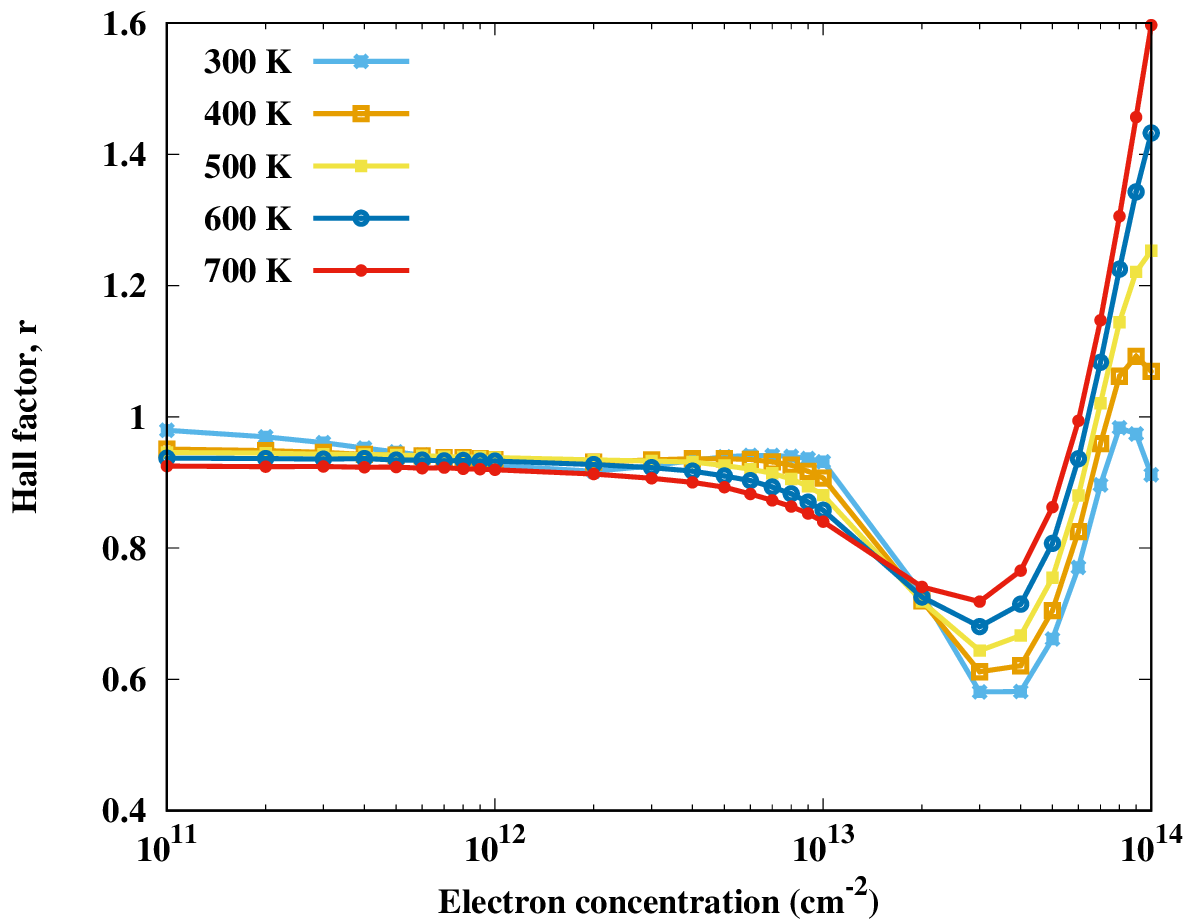}}\\
    \subfigure[$ $]{\hspace{-0.7cm}\includegraphics[scale=0.33]{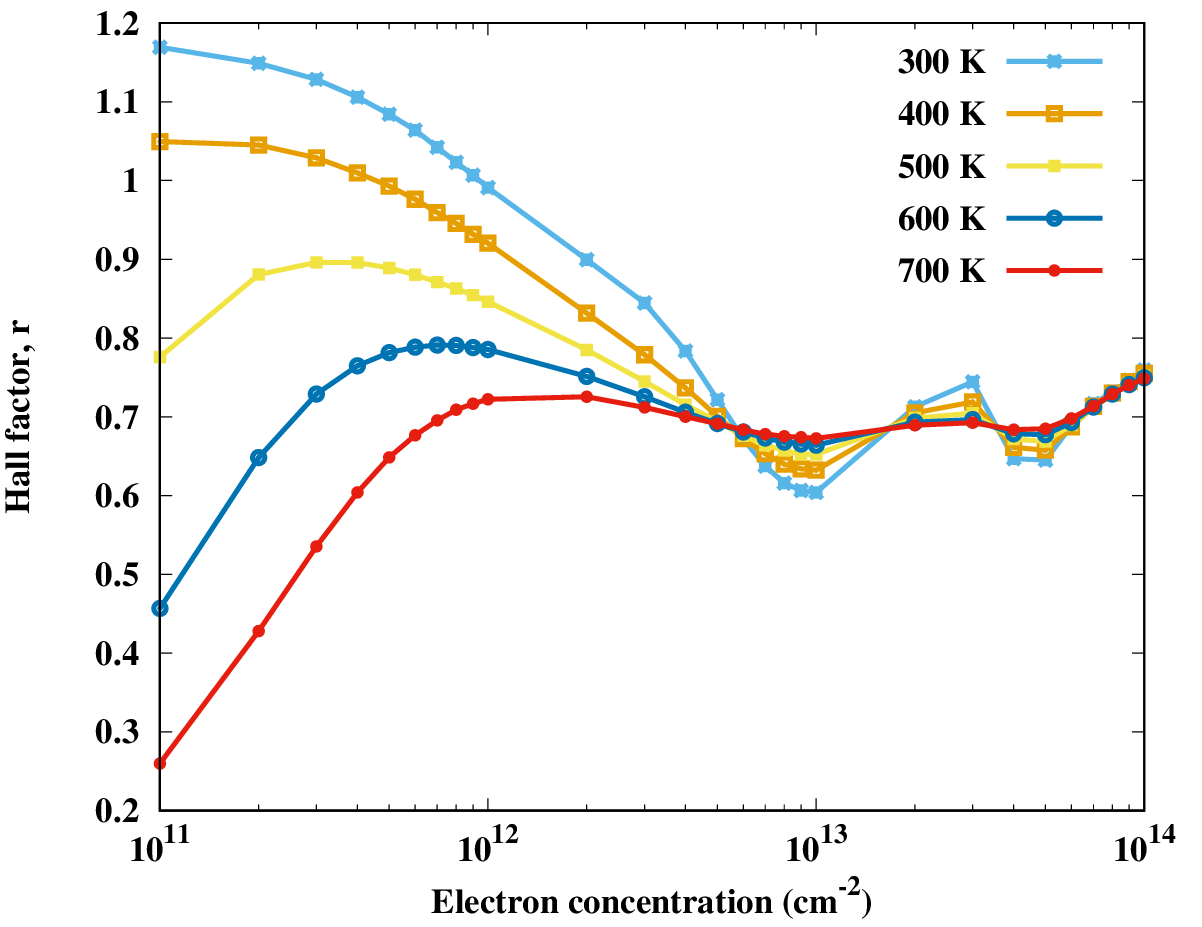}}
    \hfill
    \subfigure[$ $]{\includegraphics[scale=0.35]{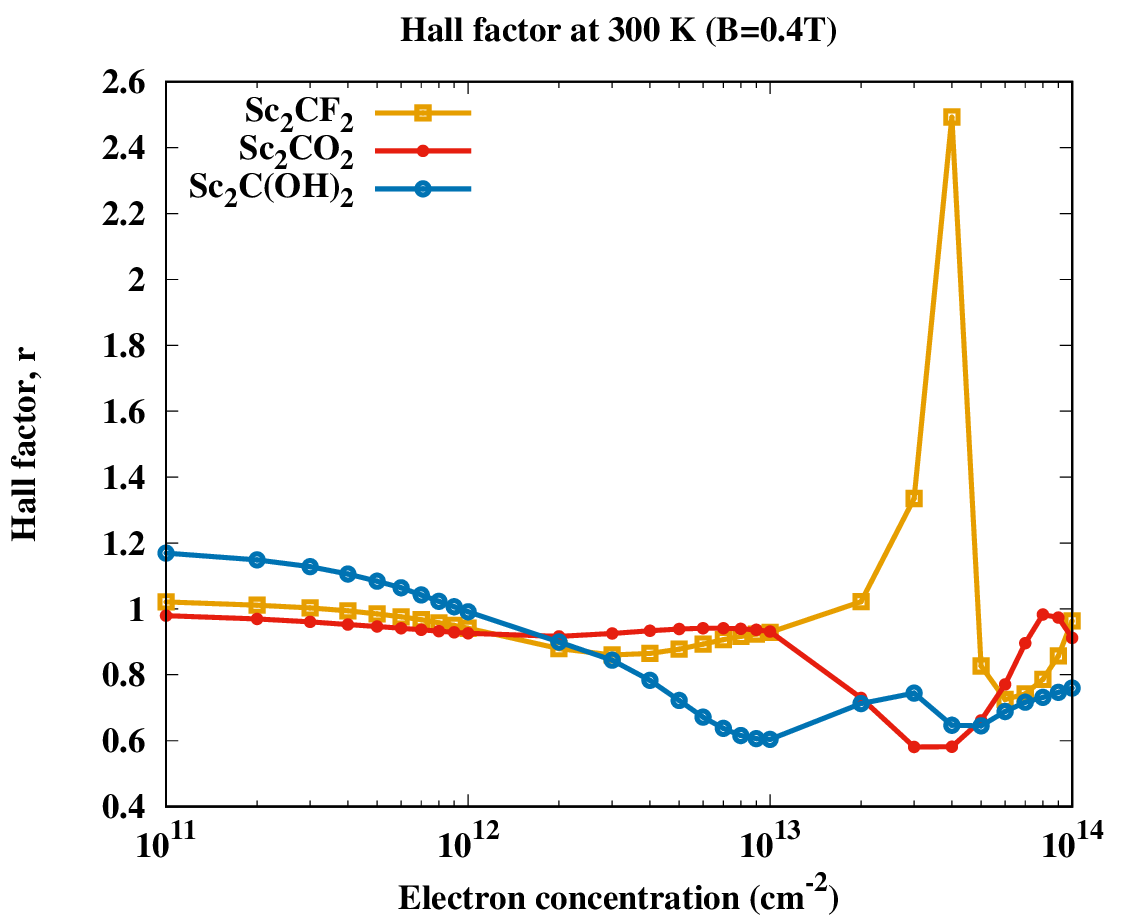}}
    \caption{The Hall scattering factor as a function of concentration at different temperatures (a)\ch{Sc2CF2}, (b)\ch{Sc2CO2} and (c)\ch{Sc2C(OH)2}. (d) Hall factor of \ch{Sc2CF2}, \ch{Sc2CO2} and \ch{Sc2C(OH)2} at 300 K.}
    \label{hall_factor}
\end{figure}
\FloatBarrier
\begin{table}
\caption{At Fermi energy (E$_{F}$) for doping concentration of 4$\times$10$^{13}$ cm$^{-2}$ and temperature of 300 K}
\begin{tabular}{ccccccccc}
Material & g(E) & h(E) & $\frac{h(E)}{((g(E))^{2}}$   \\ \\
\hline \\
\ch{Sc2CF2} & 1.63$\times$10$^{-7}$ & -5.97$\times$10$^{-12}$ & -224.69 \\
\ch{Sc2CO2} & 4.52$\times$10$^{-7}$ & -3.60$\times$10$^{-11}$ & -176.21 \\
\ch{Sc2C(OH)2} & 6.80$\times$10$^{-7}$ & -8.96$\times$10$^{-11}$ & -193.77 
\end{tabular}
\label{table1}
\end{table}
\FloatBarrier

\begin{figure}
    \centering
    \subfigure[]{\hspace{-0.7cm}\includegraphics[scale=0.33]{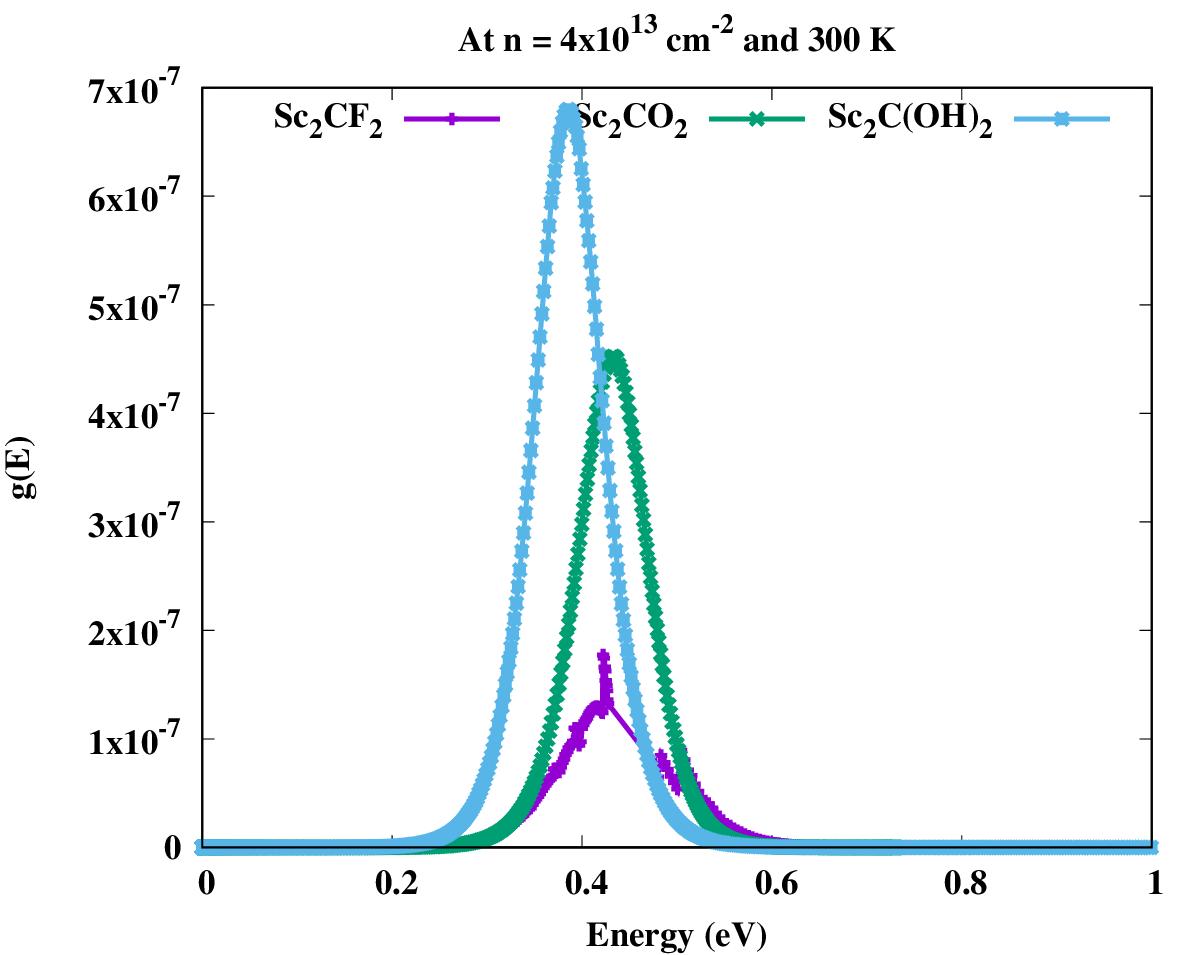}}
    \subfigure[]{\includegraphics[scale=0.33]{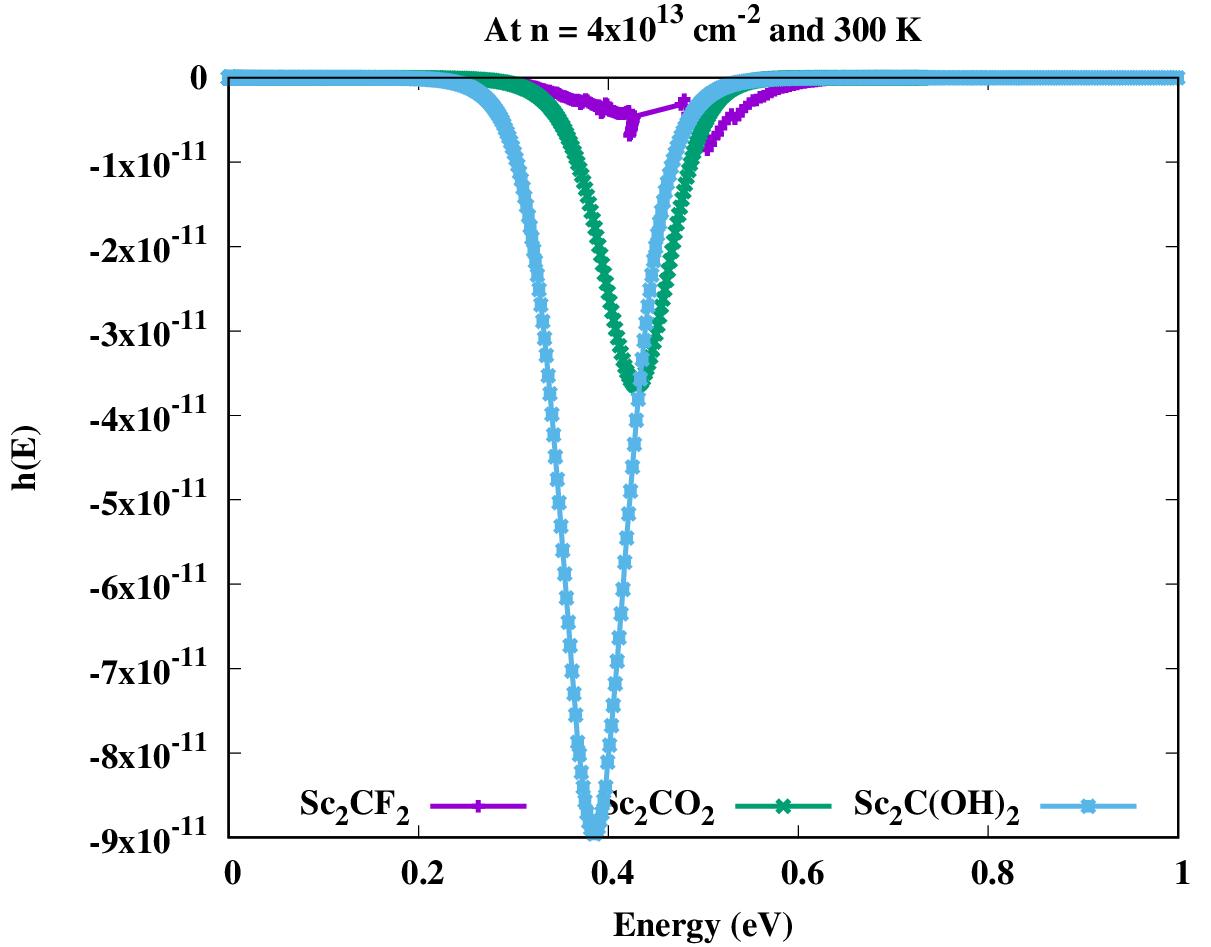}}
    \caption{The distribution functions g(E) and h(E) of \ch{Sc2CF2}, \ch{Sc2CO2} and \ch{Sc2C(OH)2} for a carrier concentration of n=4$\times$10$^{13}$ cm$^{-2}$ in the presence of magnetic field.}
    \label{g_E}
\end{figure}